\documentclass[sigconf]{acmart}

\copyrightyear{2023}
\acmYear{2023}
\setcopyright{licensedothergov}\acmConference[SIGGRAPH '23 Conference Proceedings]{Special Interest Group on Computer Graphics and Interactive Techniques Conference Conference Proceedings}{August 6--10, 2023}{Los Angeles, CA, USA}
\acmBooktitle{Special Interest Group on Computer Graphics and Interactive Techniques Conference Conference Proceedings (SIGGRAPH '23 Conference Proceedings), August 6--10, 2023, Los Angeles, CA, USA}
\acmPrice{15.00}
\acmDOI{10.1145/3588432.3591556}
\acmISBN{979-8-4007-0159-7/23/08}

\usepackage{cleveref}
\usepackage{bm}
\usepackage{amsmath}
\usepackage{float}

\usepackage{textcomp}
\usepackage{subcaption}
\usepackage{graphicx}

\usepackage{multirow}

\begin{document}



\title{Neural Face Rigging for Animating and Retargeting Facial Meshes in the Wild}

\iffalse 
    \newcommand\todo[1]{}

    \newcommand{\dafei}[1]{}
    \newcommand{\taku}[1]{}
    \newcommand{\noam}[1]{}
    \newcommand{\thibault}[1]{}
    \newcommand{\jun}[1]{}
    \newcommand{\yingruo}[1]{}
\else 
    \newcommand{\todo}[1]{{\textcolor{red}{[[TODO: {#1}]]}}}
    \newcommand{\edit}[1]{{\textcolor{red}{[[{#1}]]}}}
    \newcommand{\dafei}[1]{\textcolor{magenta}{[Dafei: {#1}]}}
    \newcommand{\taku}[1]{\textcolor{green}{[Taku: {#1}]}}
    \newcommand{\noam}[1]{\textcolor{blue}{[Noam: {#1}]}}
    \newcommand{\thibault}[1]{\textcolor{green}{[Thibault: {#1}]}}
    \newcommand{\jun}[1]{\textcolor{red}{[Jun: {#1}]}}
    \newcommand{\yingruo}[1]{\textcolor{yellow}{[Yingruo: {#1}]}}    
    \newcommand{\new}[1]{{#1}}
\fi

\def\etal{et al.~}
\definecolor{myyellow}{RGB}{252, 215, 3}
\definecolor{mygreen}{RGB}{112, 160, 65}

\author{Dafei Qin}
\email{qindafei@connect.hku.hk}
\affiliation{%
  \institution{The University of Hong Kong}
  \city{}
  \country{Hong Kong}
}

\author{Jun Saito}
\email{jsaito@adobe.com}
\affiliation{%
  \institution{Adobe Research}
  \city{Seattle}
  \country{USA}
}

\author{Noam Aigerman}
\email{aigerman@adobe.com}
\affiliation{%
  \institution{Adobe Research}
  \city{San Francisco}
  \country{USA}
}

\author{Thibault Groueix}
\email{groueix@adobe.com}
\affiliation{%
  \institution{Adobe Research}
  \city{San Francisco}
  \country{USA}
}

\author{Taku Komura}
\authornote{Corresponding author}
\email{taku@cs.hku.hk}
\affiliation{%
  \institution{The University of Hong Kong}
  \city{}
  \country{Hong Kong}
}

\begin{teaserfigure}
\centering
  \includegraphics[width=\textwidth]{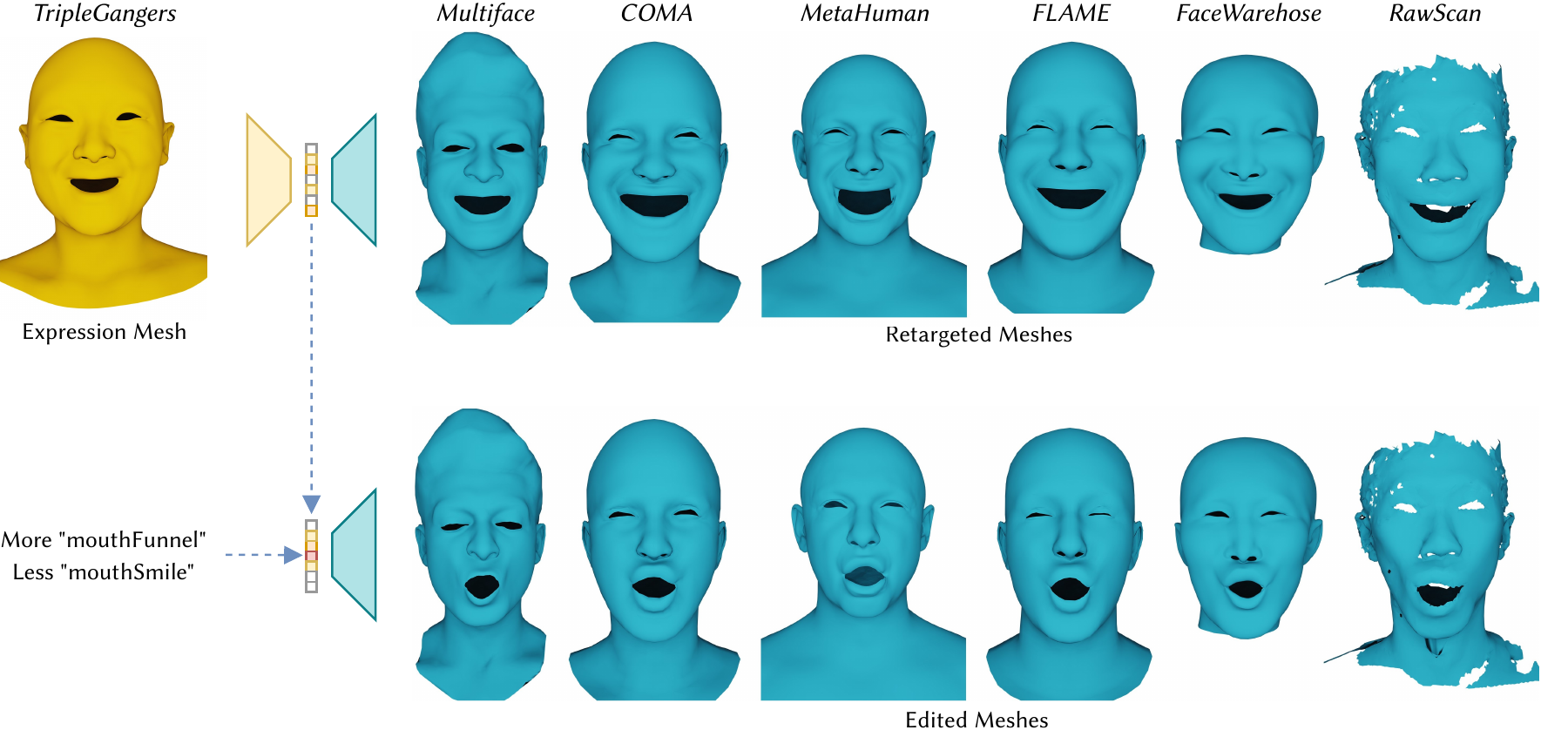}
  \caption{We present NFR, an end-to-end deep-learning approach for automatic rigging and retargeting of 3D models of human faces in the wild. \textbf{Top}: Given an unrigged facial mesh with an unknown expression and identity in an arbitrary triangulation (\textcolor{myyellow}{yellow}), NFR can transfer the expression to unrigged facial meshes with arbitrary triangulations (\textcolor{cyan}{cyan}). \textbf{Bottom}: NFR provides an interpretable latent space for user-friendly editing of the retargeted meshes. }
  \label{fig:teaser}
\end{teaserfigure}

\begin{abstract}
We propose an end-to-end deep-learning approach for automatic rigging and retargeting of 3D models of human faces in the wild. Our approach, called Neural Face Rigging (NFR), holds three key properties: 
(i) NFR's expression space maintains human-interpretable editing parameters for artistic controls;
(ii) NFR is readily applicable to arbitrary facial meshes with different connectivity and expressions;
(iii) NFR can encode and produce fine-grained details of complex expressions performed by arbitrary subjects.    
  To the best of our knowledge, NFR is the first approach to provide realistic and controllable deformations of in-the-wild facial meshes, \new{without the manual creation of blendshapes or correspondence.} We design a deformation autoencoder and train it through a multi-dataset training scheme, which benefits from the unique advantages of two data sources: a linear 3DMM with interpretable control parameters as in FACS and 4D captures of real faces with fine-grained details. Through various experiments, we show NFR's ability to automatically produce realistic and accurate facial deformations across a wide range of existing datasets and noisy facial scans in-the-wild, while providing artist-controlled, editable parameters.

\end{abstract}

\maketitle
\section{Introduction}
    This paper is concerned with leveraging deep learning for automatic rigging and retargeting of 3D meshes of human faces in the wild, supporting both raw noisy scans of real faces, as well as clean, artist-authored meshes, while exposing interpretable expression parameters which an artist can use to intuitively bring any facial mesh to a desired expression.
 
 Faces are a critical component of any human-centric tasks,  with facial movements being of significant interest in many fields of computer science including  graphics, computer vision, and HCI. 
 Many computational representations of 3D human faces have been proposed,
 with early formulations such as \emph{Facial Action Coding System } (FACS)~\cite{Ekman1978FacialAC} and PCA-based learned linear \emph{3D Morphable Models} (3DMM)~\cite{blanz1999morphable}. However, these rigged models can only operate on faces with known geometry and triangulation (where a consistent correspondence is provided), and cannot be automatically extended to novel face models with arbitrary triangulations.

Thus, the goal of this paper is to provide a method to understand and control real-world facial data with the following properties:
\begin{itemize}

\item  The deformation can be controlled via \emph{interpretable} parameters, enabling users to intuitively control expressions.
\item  The method should be applicable to arbitrary human faces of unknown subjects and expressions. It should generalize to arbitrary triangulations, and be robust to both noise and missing areas on the 3D face.

\item  The method should produce highly-accurate expressions. It should accurately encode and decode the fine-grained details of nonlinear deformations on real human faces.
\end{itemize}
    
We address these challenges by developing an autoencoder for facial expressions. Specifically, we design an encoder to transform a given face mesh into a latent code representing its expression. We then employ recent advancements in deep learning of 3D deformations~\cite{10.1145/3528223.3530141} to construct a decoder that takes an expression code and alters the face to the desired expression. To ensure the capture of all facial features and expression nuances, we create an encoder that combines image encoding of input face renderings with a state-of-the-art mesh encoder~\cite{sharp2022diffusionnet}. Our encoder features two branches—one for input expression and one for facial identity—and incorporates a training setup that separates identity from expression.

 Lastly, we aim to learn the \emph{interpretable} latent space, i.e., that each entry in the code can be interpreted as a parameter corresponding to a specific, meaningful localized activation on the human face. This goal faces a significant challenge posed by the lack of data: existing datasets comprise real human scans which are not directly coupled with expression parameters, hence we need to devise an architecture and a training scheme that will disentangle interpretable parameters from human scans with unknown expressions.

Specifically, we use two different complementary datasets to achieve our goal: 1) \emph{Multiface}~\cite{wuu2022multiface} which is rich in natural and identity-specific deformations, but lacks interpretable expressions; 2)  \emph{ICT FaceKit}~\cite{li2020learning}, which has interpretable parameters as in FACS, however, the expressions are synthetic and less realistic. We then use these two datasets in a training scheme of our expression autoencoder. We train the system such that for the ICT dataset the latent code matches the ICT deformations parameters,
while for the \emph{Multiface} dataset, the rich and expressive details of the faces are reconstructed. 
This joint training empirically leads to a latent space that directly matches the FACS parameters, while further representing the rich deformations in the \emph{Multiface} dataset.
Our framework thus couples the interpretable FACS parameters with the capability to produce realistic deformations as in the \emph{Multiface} dataset.

We show through experiments that our framework is able to perform various tasks, such as deformation transfer from unknown faces to other faces, as well as user editing of facial expressions. We show our network carries its abilities across various models, from artist-authored artistic meshes up to noisy, partial face scans in the wild.
Our system significantly improves the efficiency of the human facial animation pipeline, by-passing heavy data pre-processing such as facial alignment, remeshing, and manual rigging.

\section{Related Works}
    In this section, we review techniques about facial deformation models and deformation transfer.  
\subsection{Facial Deformation Models}
\paragraph{Anatomy-inspired models}
Facial Action Coding System (FACS) \cite{Ekman1978FacialAC} defines facial movements as the combination of muscle activations, or \emph{Action Units} (AUs). Variants of FACS have been adopted in graphics and animation for their intuitive artistic controls, typically implemented with blendshape deformers~\cite{lewis2014practice}. Such FACS models can be viewed as 3D Morphable Models (3DMM)~\cite{blanz1999morphable} with hand-crafted basis using the domain knowledge of the human anatomy. This in turn means FACS-based 3DMMs require manual sculpting of many shapes. The inverse rig problem (a.k.a. retargeting in the context of facial animation) which solves the optimal 3DMM parameters fitting to target shapes is also not trivial for conventional FACS-based models~\cite{lewis2010direct, seol2011artist, cetinaslan2020sketching, cetinaslan2020stabilized}. 

Several recent studies learn compact and sparse neural representations from FACS models. Bailey et al.~\shortcite{bailey2020fast} replaces film-quality animation rigs with a learned deep model. \new{Vesdapunt et al.~\shortcite{vesdapunt2020jnr} propose a person-specific joint-based neural skinning model with highly compact and sparse latent space. 
Choi et al.~\shortcite{choi2022animatomy} design an intuitive interface for controlling character expressions through curves drawn on the face along facial muscle structures.  
Although these models preserve intuitive deformation controls, they are subject-specific, i.e., building these models for a new character requires starting from scratch. Our goal is to maintain the intuition of FACS but learn subject-specific rigging and retargeting from data with minimal manual labor.}
\paragraph{Data-driven models}
To avoid the ad-hoc manual sculpting of 3DMM basis, learning linear 3DMMs from face scans is a popular choice~\cite{blanz1999morphable, li2017learning, paysan2009, HuberP2016AM3M, choe2001performance, choe2006analysis, wu2016anatomically, tewari2017mofa, brunton2014multilinear}.
The readers are referred to~\cite{3DMM_survey} for a thorough survey of 3DMM related methods. Limited by the linear nature, these models struggle to produce highly variant and complex facial expressions. \new{Additionally, the PCA bases fail to provide interpretable and sparse controllers.} 

Mesh-based neural networks are proposed to learn more expressive facial representations than linear 3DMMs. Monti et al.~\shortcite{monti2017geometric} propose a unified framework to operate on 3D mesh local geometries.  Ranjan et al.~\shortcite{ranjan2018generating} apply spectral convolutions on 3D meshes and uses hierarchical sampling to capture local and global features. Verma et al.~\shortcite{verma2018feastnet} enable the model to dynamically learn the correspondence between the data during training. Gong et al.~\shortcite{gong2019spiralnet++} propose spiral convolution to retrieve information on triangle mesh neighborhood. Bouritsas et al.~\shortcite{bouritsas2019neural} follow a similar approach to directly processing vertex offsets and achieves better results on face representation than 3DMM.  \new{Song et al.~\shortcite{song2020accurate} propose to learn the facial model in the differential subspace. Zhou et al.~\shortcite{zhou2020fully} generalize the 3D mesh autoencoder to train on tetrahedra and non-manifold meshes.  Because these methods depend on specific mesh templates, they fail to apply to meshes with different representations.} 

\new{Several recent works target triangulation agnosticism.} \new{Chandran et al.~\shortcite{chandran2022shape} utilize a transformer architecture and positional encoding to align input meshes to a canonical space. While supporting different triangulations, correspondence is required if users want to apply their model to unseen facial meshes.}   \new{Yang et al.~\shortcite{yang2022implicit} propose a physically-based implicit model to control soft bodies like human faces, which supports different mesh resolutions of the same identity.} 

In summary, meshes from one face dataset/model cannot be repurposed for other datasets/models without going through intensive processing steps using non-rigid registration guided by manual landmarks to take the dense correspondence~\cite{li2020learning}. Our method removes this burden by training  on multiple datasets with different representations, enabling application to diverse facial meshes.

\begin{figure*}[ht]
  \centering
    \includegraphics[width=0.9\linewidth]{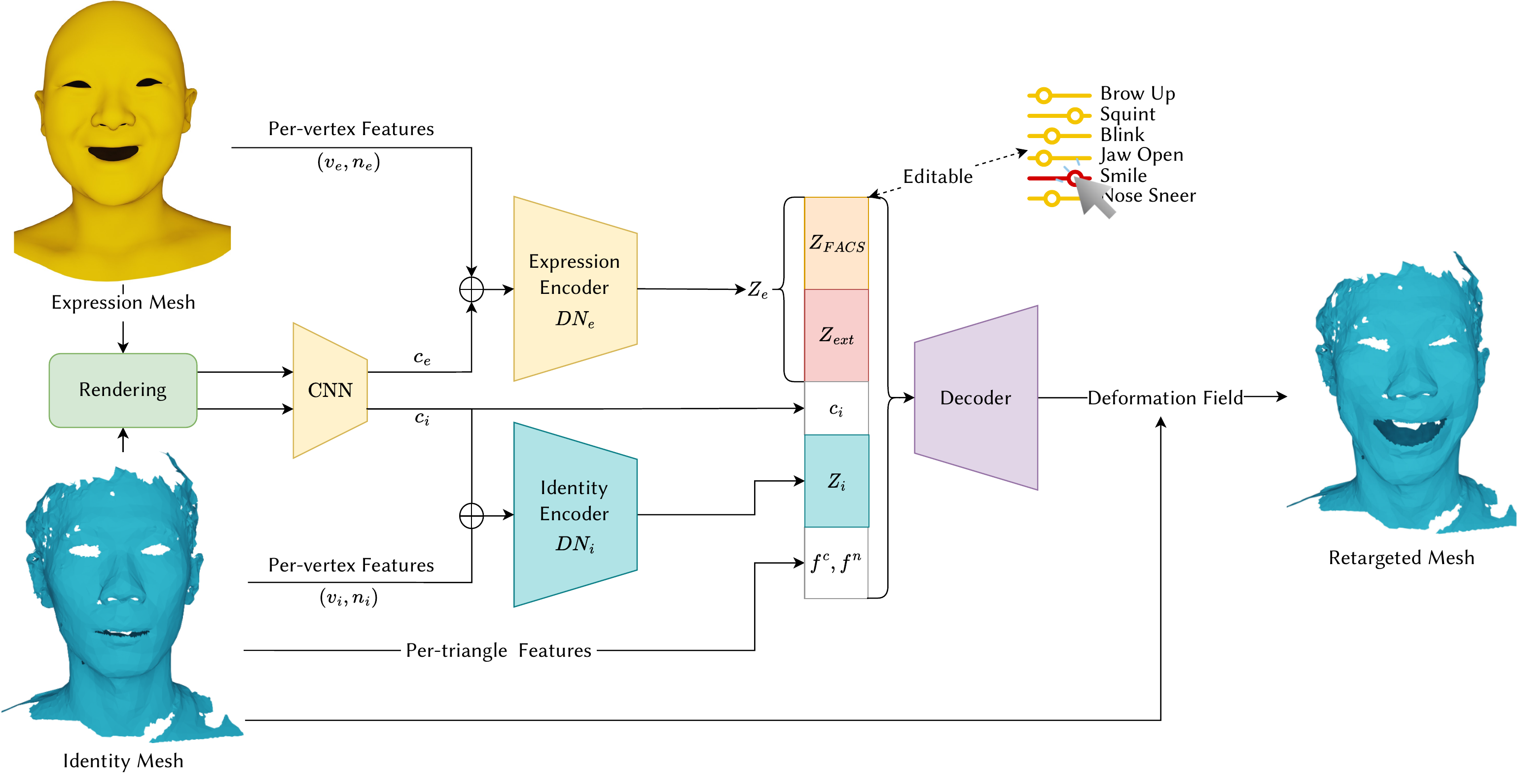}

  \caption{\textbf{Overview.} Given an unrigged facial mesh with an unknown expression and identity in an arbitrary triangulation (\textcolor{myyellow}{yellow}) and another unrigged \textit{neutral} facial mesh with the target identity in an arbitrary triangulation (\textcolor{cyan}{cyan}), NFR  extracts an expression code $z_e$ from the first mesh, and an identity code  $z_i$ in the second mesh, and transfer the expression to the target identity with non-linear deformations as in real human faces. A part of $z_e$, denoted as $z_{FACS}$, follows the interpretable rigging parameters, allowing NFR to behave as an artist-friendly auto-rigger and retargeter.}

  \label{Fig:overview}
 \end{figure*}

\subsection{Deformation Transfer}

\new{\textit{Deformation Transfer} is a retargeting technique that directly works on meshes. Sumner and Popović~\shortcite{sumner2004deformation} transfer the animation by mapping the deformation gradients from the source to the target. Li et al.~\shortcite{li2010example} propose to build a target blendshape model by combining existing rigging prior and a few target example expressions.  These methods require dense correspondence of the source-target pair. }

\paragraph{\new{Neural Deformation Transfer}}

The investigation of object deformation through neural networks~\cite{gao2018automatic} have been extensively explored. A comprehensive overview can be found in the survey by \cite{deformationsurvey}.
Tan \etal\shortcite{Tan_2018_CVPR} establish a mesh VAE for learning deformation spaces of a specifically given mesh. Gao \etal\shortcite{gao2018automatic} establish automatic deformation transfer between two mesh datasets, which does not require explicit correspondence between the pair of meshes. However, these methods assume the input mesh structures are consistent:  Mesh VAEs cannot aggregate information from multiple datasets where the connectivities of the meshes are different. Retraining is required for transferring the deformation between new mesh pairs. 

\new{Certain facial models~\cite{jiang2019disentangled, chandran2020semantic, chandran2022facialanimation} disentangle identity and expression spaces, simplifying deformation transfer within the identity space. Still, these models do not accommodate mesh templates divergent from the training set, precluding transfer to or from custom meshes. Moser \etal\shortcite{moser2021semi} suggest transferring animations between rendered videos and 3D characters via an image-to-image model, but this approach is character-specific and sacrifices deformation interpretability with PCA bases.}
Notably, none of these methods addresses the primary objective of this paper: integrating the neural deformation space with interpretable parameters to facilitate fine-grained human control over deformation while accommodating in-the-wild facial meshes.

\section{Method}
    
In this section, we first give an overview of the NFR framework. Then we explain how to leverage existing face datasets and 3DMMs to train a network that is both interpretable and generalizable to meshes with various shapes and triangulations. 

\subsection{Architecture}

In essence, the architecture of NFR can be described as that of an autoencoder (Fig. \ref{Fig:overview}).
Specifically, given a neutral identity mesh $M_i$ to be deformed to an expression, and an expression mesh $M_e$ representing the desired expressions, the expression encoder first maps $M_e$'s deformation into a FACS-like latent expression code $z_e$. Similarly, the identity encoder maps $M_i$ into an identity code $z_i$.  Based on $z_e$ and $z_i$, the decoder deforms $M_i$ to $M_e^*$ to approximate the expression mesh as best as possible $M_e^*\approx M_e$. The trained deformation decoder thus acts as a high-fidelity facial rig applicable to any facial mesh, with human-friendly controls (its FACS-like latent space). The end-to-end pipeline allows NFR to automatically transfer facial expressions to different identities while maintaining interpretability.
To accomplish this, we leverage the recent advances in triangulation-agnostic neural geometry learning. Namely, we use a combination of image CNN's applied to renderings, along with \emph{DiffusionNet} (DN)~\cite{sharp2022diffusionnet} to build encoders to extract 3D shape features, and use \emph{Neural Jacobian Fields} (NJF)~\cite{10.1145/3528223.3530141} as a decoder to  produce deformations of the facial meshes. 

\paragraph{Identity Encoder}
We first feed a front-view rendering of $M_i$ through $CNN$, a plain 2D CNN, to receive a code $c_i \in \mathbb{R}^{128}$. This $c_i$ is then fed into $DN_i$, a DiffusionNet-based encoder. DiffusionNet is applied on \textit{Per-vertex features}, which is the local shape features of $M_i$, namely the concatenation of the vertex coordinates $v_i$ and vertex normals $n_i$. 
Intuitively, combining the 2D CNN and the Per-vertex features enables $DN_i$ to capture subtle facial features. $DN_i$ \new{receives the neutral identity as inputs}, and outputs an identity latent code $z_i \in \mathbb{R}^{100}$, which is the weighted means of the per-vertex output.

\paragraph{Expression Encoder} Similarly, for the expression encoder, the rendering of $M_e$ is fed into $CNN$ to produce a code $c_e \in \mathbb{R}^{128}$, which is concantenated with the \textit{Per-vertex features}, $(v_e, n_e)$ to be fed to a DiffusionNet-based encoder $DN_e$. In contrast to $DN_i$, $DN_e$ \new{receives a face mesh with an expression}, and outputs a predicted expression code $z_e \in \mathbb{R}^{128}$ whose elements are trained to be control parameters that are interpretable. During training, the first 53 dimensions of $z_e$ are regularized to mimic the ARKit-compatible FACS. We denote them as $z_{FACS}$. The remaining part of $z_e$, denoted as $z_{ext}$, is learned from the data. This explicit decoupling is critical for NFR to learn realistic expressions from real scans while maintaining maximum interpretability for editing.

\paragraph{Decoder} We adopt NJF as the decoder, where the core is a multi-layer perception (MLP). The MLP takes \textit{Per-triangle features}, i.e. triangle centers $f_i^c$ and normals $f_i^n$ of the identity mesh $M_i$, the expression code $z_e$, the identity code $z_i$ and the shape code of the identity mesh $c_i$ as the input, and outputs the deformation Jacobian $g^* \in \mathbb{R}^{3\times3}$ for each triangle.

\subsection{Datasets}

The crux of our approach lies in the simultaneous training of both actual facial expressions from scanned faces and computer-generated expressions created through the manipulation of parametric controls on rigged meshes.
As a result of this strategy, our neural method is capable of generating highly realistic expressions for these parameterized controls.
Hence, we leverage two types of data: synthetic data and real scans.

\paragraph{Synthetic data.} \emph{ICT FaceKit (ICT)}~\cite{li2020learning} is a linear 3DMM for facial expressions. It has 153 control parameters: a 100-dimensional PCA-based identity space, and a 53-dimensional FACS-inspired hand-crafted expression rig. This FACS model is compatible with the Apple ARKit blendshape model~\footnote{https://developer.apple.com/documentation/arkit/arfaceanchor/blendshapelocation} which allows us to collect data with plausible AU activations with ARKit face tracking. These AUs have pre-defined semantics, \textit{e.g}, the first parameter, which is called 'browInnerUp\_L',  corresponds to lifting the left inner eyebrow.
We adopt a template of ICT that has 3,694 vertices and 7,007 triangles after the mesh standardization (Sec. \ref{subsec:data-augmentation}).

We generate two synthetic datasets with \emph{ICT FaceKit} 
\begin{itemize}
    \item  \emph{ICT-Random-AU}. We sample eight random identities and 5,819 random AUs per identity.  \emph{ICT-Random-AU} thus contains 41,322 samples. We select the first six identities as the training set, one as validation, and one as testing. 
    \item \emph{ICT-Real-AU}. We sample 22 random identities from ICT and 1,630 frames from ARKit face tracking of our own performances to capture plausible AU activations. We perform an 8:1:1 split on the frames. \emph{ICT-Real-AU} thus contains 32,600 training frames, 3260 validation frames, and 3260 test frames.
\end{itemize}

\paragraph{Real Scans.} \emph{Multiface}~\cite{wuu2022multiface} is a large facial dataset with fine-grained details from 4D scans. It contains 13 different identities. Each subject is captured by a multi-camera setup at 30 FPS while reading a script designed to cover over a hundred facial expressions. We use the same train/test split as in~\cite{wuu2022multiface} while omitting expressions that are too close to the neutral face from the training set. The training split contains 28,991 meshes and the testing split contains 13,610 meshes. Each mesh has 5,385 vertices and 10,581 faces after applying the mesh standardization.

\subsection{Data Augmentation and Standardization}
\label{subsec:data-augmentation}
\paragraph{Augmentation.} In-the-wild facial meshes vary in many ways: some have no necks, some have no back of the head, and some have bad geometry regions and holes. We conduct two types of data augmentation to adjust to these scenarios. (i) randomly shift the input off the center and scale independently along the $x, y,$ and $z$ axis, and apply masks that remove parts of the face.  (ii) randomly cut holes on the input mesh to provide geometrical variance. 
We will show in Sec.\ref{subsec:Ablations} this data augmentation is crucial for generalizing to in-the-wild  inputs.

\paragraph{Standardization} Facial datasets have different ways of modeling the internal structures of human faces, most notably eye sockets and the oral cavity. E.g. \emph{ICT} has realistic eye sockets, whereas \emph{Multiface} covers up eye openings with meshes. Therefore, we cut out the eye and mouth internals from mesh templates to standardize.

\subsection{Multi-dataset training}

We introduce a multi-dataset training scheme to aggregate information from the synthetic data  and real scans. We first train on data generated from \emph{ICT FaceKit} to initialize our model to imitate a linear 3DMM, then we further train on a mix of real scans from  the \emph{Multiface} dataset and synthetic data. 

\paragraph{Decoder Training}
Throughout the training, we use the same loss on the decoder, which contains a vertex $L_2$ loss: $L_v = ||v_e - v_{e^*}||^2$, a Jacobian loss $L_g = ||g - g^*||^2$ and a normal vector loss $L_n = ||n_e - n_e^*||^2$. Here $g$ is the ground truth deformation Jacobian from $M_i$ to $M_e$. $v_e^*$, $n_e^*, g^*$ are the vertex positions, vertex normals, and deformation Jacobians of the output deformed mesh $M_e^*$. Given the weights of three loss terms $\lambda_v$, $\lambda_g$ and $\lambda_n$, the total loss of the decoder is defined as:
\begin{equation}
    L_{dec} = \lambda_v L_v + \lambda_g L_g + \lambda_n L_n.
    \label{eq:recontruction}
\end{equation}

\paragraph{Training on ICT FaceKit.} We first train the model with the two \emph{ICT} datasets and explicitly supervise $z_{FACS}$ with an $L_2$ loss to reproduce the same expression code as \emph{ICT}. The remaining latent expression code, $z_{ext}$, is forced to be zero. We warm up the decoder by feeding the ground truth $z_e$ to the MLP. The model is thus initialized to imitate a linear 3DMM. We define the output expression code as $z_e^* = [z_{FACS}^*, z_{ext}^*]$. The encoder loss is defined as:
\begin{equation}
    L_{enc} = ||z_{FACS}^{ } - z_{FACS}^*||^2  + ||z_{ext}^*||^2.
\end{equation}
 The total loss is a weighted sum of $L_{enc}$ and $L_{dec}$: $L = \lambda_e L_{enc} + L_{dec}$.

\paragraph{Training on Multiface and ICT FaceKit} Second, we train with \emph{Multiface} to strengthen the network's ability to represent fine-grained expressions that out-performs linear 3DMM. The challenge is that \emph{Multiface} does not have ground truth latent expression parameters. To maintain an interpretable latent space, we rely on the fact that NFR is correctly initialized to form a FACS-like latent space. We keep a part of the batch from \emph{ICT-Real-AU} with direct latent supervision to maintain this interpretability.   \emph{ICT-Random-AU} is not used here since this randomly generated dataset may contain unrealistic expressions. On \emph{Multiface} we simply regularize the latent space if the parameters go out of the range $[0, 1]$ to follow the convention of FACS AUs:
\begin{equation}
L_r(x) = \left\{\begin{aligned} -x&,& x < 0 \\
                0&, &0 \le x \le 1 \\
                    x - 1&. &x > 1 \end{aligned}\right. 
\end{equation}

The loss of the encoder at this stage becomes:
\begin{equation}
    L_{enc} = \left\{\begin{aligned}
         ||z_{FACS}^{ } - z_{FACS}^*||^2  + ||z_{ext}^*||^2&, & \emph{ICT-Real-AU}\\
         L_r(z_e^*)&. &\emph{Multiface}
    \end{aligned}
    \right.
\end{equation}

After these two stages of training, our model can retarget realistic expressions to in-the-wild meshes and preserves an interpretable latent expression space for manipulation.

\section{Experiments}
    In this section, a series of experiments are conducted to showcase the capabilities of NFR. Sec.~\ref{subsection: InverseRigging} evaluates the encoder's effectiveness by comparing inverse-rigging outcomes 
 for \emph{ICT-Real-AU} against \emph{Seol} \cite{seol2011artist}. Sec.~\ref{subsec:reconstruction} benchmarks the expression quality of NFR against template-specific mesh reconstruction methods. Sec.~\ref{subsec:tri-agnostic} confirms the triangulation-agnostic property through inverse rigging experiments on \emph{ICT-Real-AU} with a different mesh template.  Sec.~\ref{subsec:retargeting} demonstrates NFR's practicality by retargeting expressions to in-the-wild meshes of varying connectivity and resolutions.  Sec.~\ref{subsec:editing} further illustrates the model's practicality, plotting the activations of each position in $z_{FACS}$ to reveal a semantically related latent expression space. The ease of use is demonstrated through two sequences of editing processes applied to in-the-wild expressions. Sec.~\ref{subsec:non-linear} qualitatively exhibits that sampling in the learned expression space produces plausible expressions, surpassing a basic 3DMM. Lastly, Sec.~\ref{subsec:Ablations} presents various ablation studies to substantiate the key design choices of the model and training scheme.
The readers are referred to the supplementary video for the details.  

 We use the per-vertex Euclidean distance as the error metric and additionally report the 90\% percentile value to emphasize the bad cases.

\subsection{Inverse Rigging}
\label{subsection: InverseRigging}

\emph{Inverse rigging} is a task to find optimal rigging parameters fitting the resulting deformation to a given  geometry. 
Our encoder behaves as a triangulation-agnostic inverse rig predictor extracting FACS AUs from a facial mesh with unknown, entangled AU activations.

Table.~\ref{tab:table-ict-live} compares the inverse rigging results of our method against \emph{Seol} \cite{seol2011artist}, an iterative optimization-based method that relies on the ground truth rigging model. Here, \emph{Ours (ICT rig)} represents the deformation from ICT's linear 3DMM driven by our inverse-rigged parameters. \emph{Ours (NFR)} represents the deformation output of our decoder. In this complex setting where many AUs are activated simultaneously, our method outperforms \emph{Seol} by a large margin, where the performance of \emph{Ours (ICT rig)} validates the effectiveness of the expression encoder, and that of \emph{Ours (NFR)} indicates that the rigging space of our model is close to the ICT rig. 

\begin{figure}
     \centering
     \includegraphics[width=.9\linewidth]{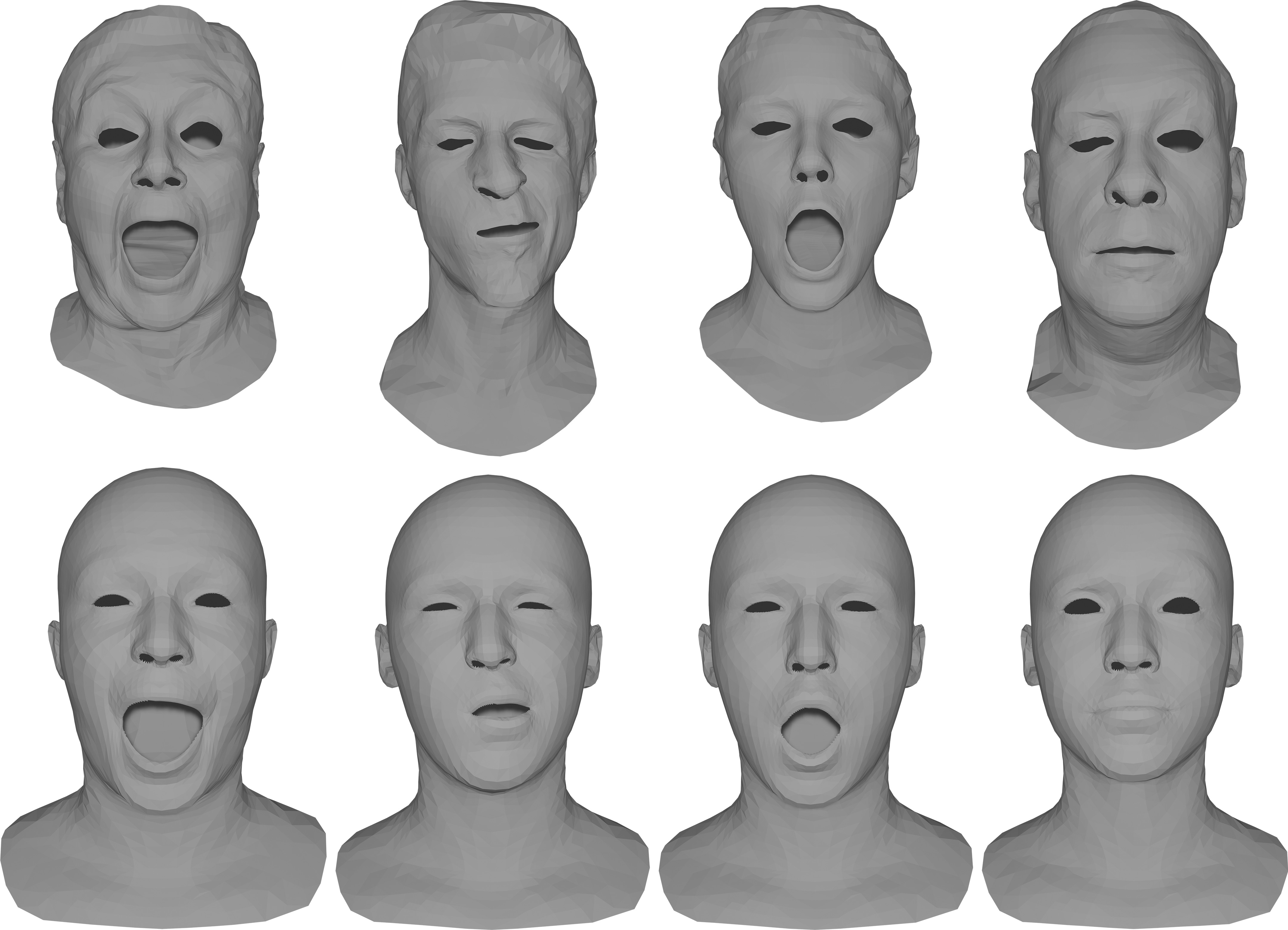}
     \caption{\textbf{Inverse Rigging.} \textbf{Top}: The input \emph{Multiface} expressions. \textbf{Bottom}: Applying the latent $z_{e, FACS}$ to the ICT rig. All deformations on the eyebrows, eyes, and mouth are solved correctly.}
     \label{fig:mf-inv-rig}
 \end{figure}

In Fig.~\ref{fig:mf-inv-rig} we apply $z_{FACS}$ of \emph{Multiface} on the ICT rig. Though limited by the expression ability of the linear rig, the expressions on the ICT rig are still semantically close to that of \emph{Multiface}. This validates the inverse rigging ability of NFR on in-the-wild expressions. Note that while \emph{Ours (NFR)} performs slightly worse on the synthetic dataset \textit{ICT-Real-AU} than \emph{Ours (ICT rig)}, NFR's major advantage is to work on in-the-wild data with arbitrary triangulation and high-frequency details.

\begin{table}\footnotesize
\centering
 \caption{
 \label{tab:table-ict-live}
 \textbf{Inverse Rigging on \emph{ICT-Real-AU}}. Our encoder, evaluated both via \textit{ICT rig} and \textit{NFR}, outperform \emph{Seol}~\cite{seol2011artist} by a large margin.}
\begin{tabular}{ lcccc } 
 \toprule
Method & Mean (\textit{mm})   & Median  (\textit{mm}) & 90\%  (\textit{mm})\\
 
 \midrule
 Seol~\cite{seol2011artist} & 0.688 $\pm$ 0.979 & 0.369 & 1.640\\

 Ours (ICT rig) & \textbf{0.378 $\pm$ 0.467} & \textbf{0.225} & \textbf{0.860}\\ 
 Ours (NFR) & 0.443 $\pm$ 0.454 & 0.307 & 0.920\\
 \bottomrule
\end{tabular}

 \end{table}  

\subsection{Expression Quality}
\label{subsec:reconstruction}
In this section, we assess the ability of our model to generate precise geometric details. The expression encoder is employed to map target deformations into an interpretable expression space, subsequently reconstructing the input deformation through the decoder. Our model is compared to three template-specific competitors: COMA~\cite{ranjan2018generating}, N3DMM~\cite{bouritsas2019neural}, and SpiralNet++~\cite{gong2019spiralnet++}. All models are trained on the same \emph{Multiface} training split, with our model also receiving training on synthetic \emph{ICT} datasets. \new{NFR is designed to maintain an interpretable latent space and handle meshes in a triangulation-agnostic manner, whereas competitor models are constrained by a single mesh template, limiting their capacity to learn from diverse datasets and resulting in less interpretability. latent spaces. A baseline method, \emph{NFR (Multiface)}, trained solely on the \emph{Multiface} dataset, demonstrates the advantages of multi-dataset training. }

\begin{figure}[h!]
  \centering
    \includegraphics[width=0.9\linewidth]{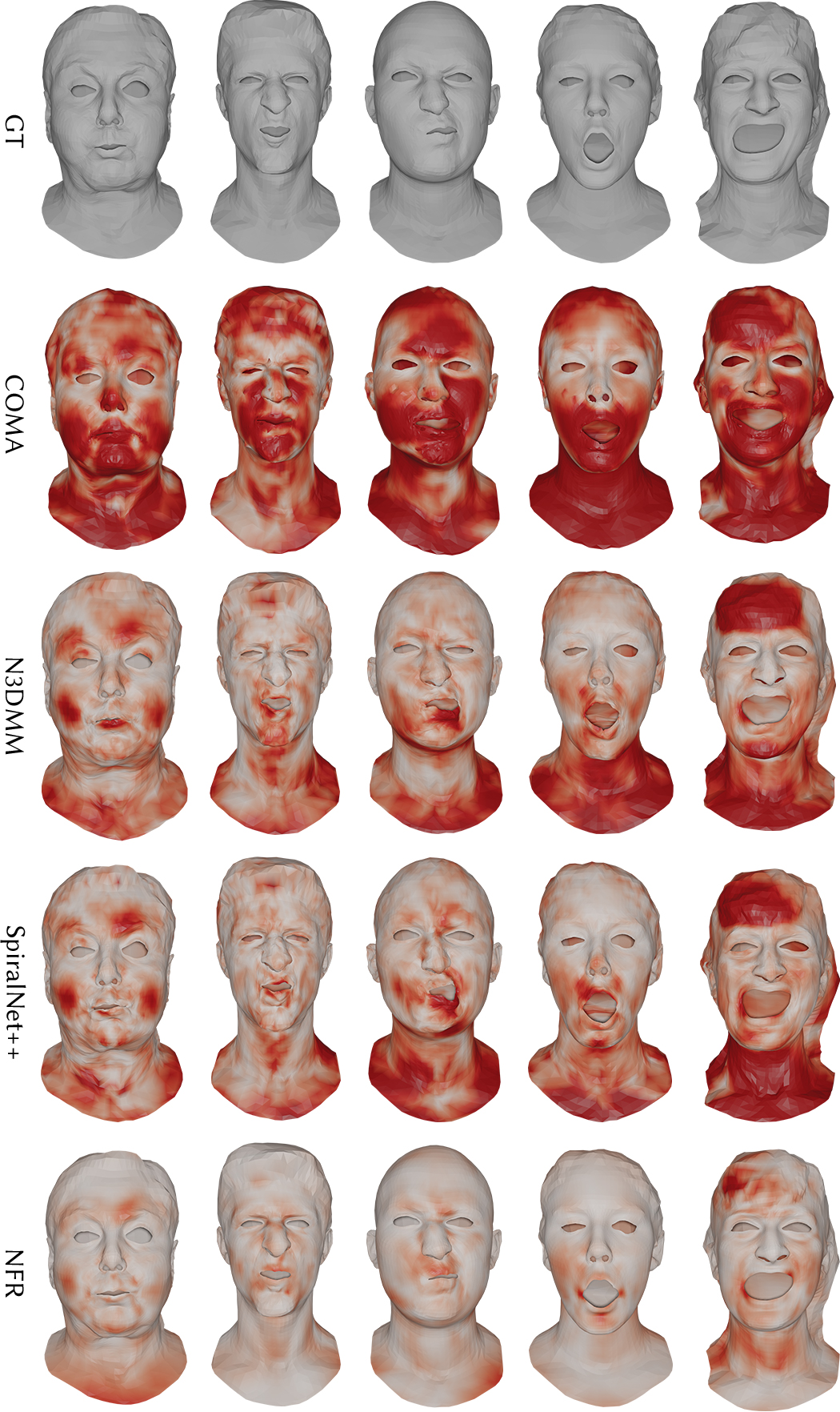}
  \caption{\textbf{Qualitative reconstruction results on \textit{Multiface}}. We color the meshes by their per-vertex Euclidean error. We compare COMA, SpiralNet++, N3DMM, and NFR. The color map clearly shows that NFR outperforms the competing approaches.}
  \label{Fig:MF_colored}
 \end{figure}

\begin{figure}[h!]
  \centering
    \includegraphics[width=0.9\linewidth]{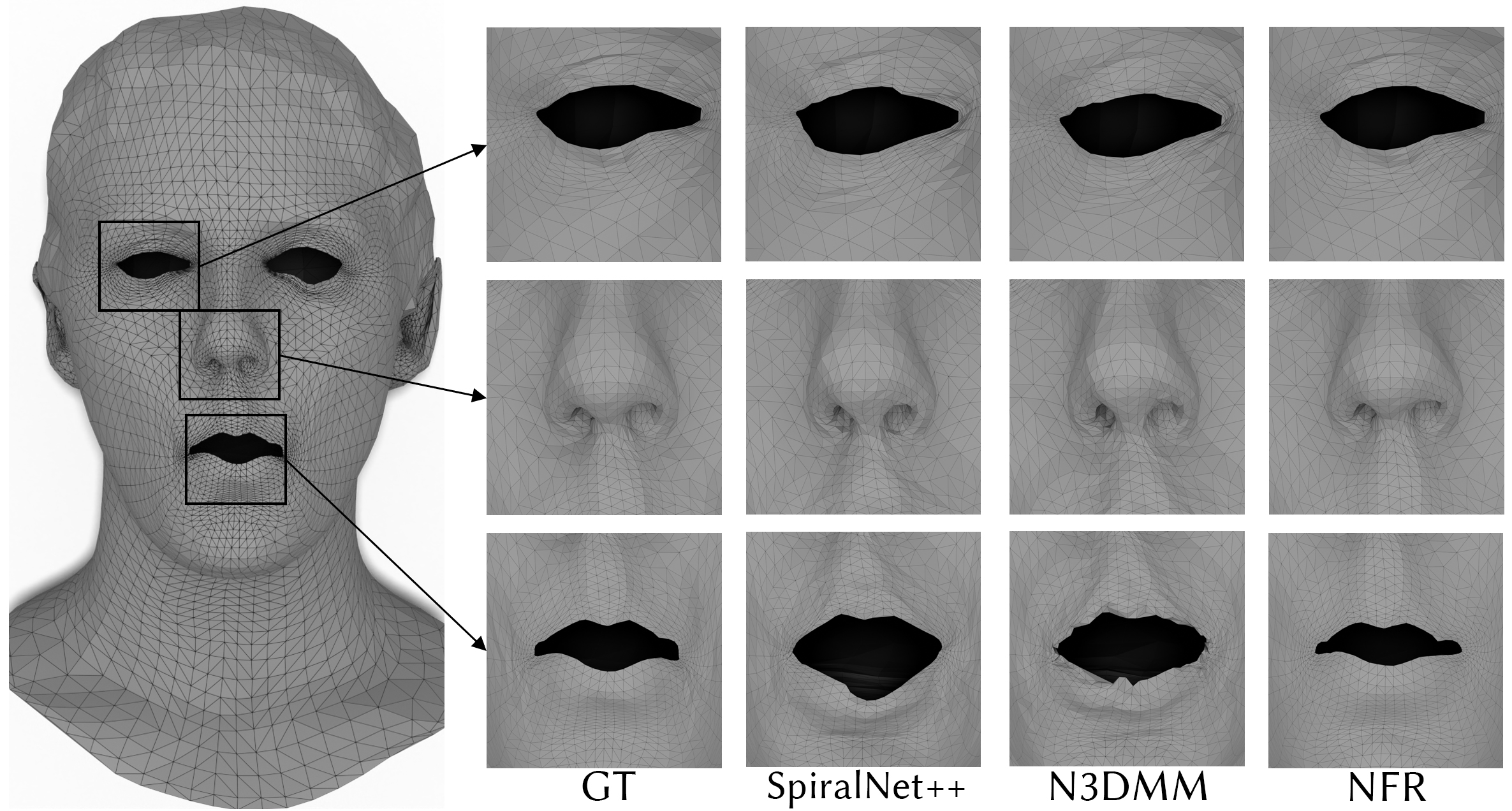}
  \caption{\textbf{Zoomed-in comparison on ~\textit{Multiface}}. NFR not only gets rid of the coarse surface artifacts of the other baselines but also preserves the shape better around the eyes, nose, and mouth. Left is the ground truth.}
  \label{Fig:MF_detailed}
 \end{figure}

Table.~\ref{table-mf} reveals that NFR significantly outperforms its competitors, achieving a 32.3\% reduction in the 90\% percentile value compared to the best-performing competitor. \new{Training on multiple datasets establishes a sparse and semantic latent space and enhances expression reconstruction.} In Fig.~\ref{Fig:MF_colored}, we visualize the per-vertex Euclidean error on \emph{Multiface} samples and provide a magnified comparison in Fig.~\ref{Fig:MF_detailed}. NFR exhibits fewer coarse surface artifacts than other baselines and more effectively preserves the shape around facial features such as the eyes, nose, and mouth.
 
\begin{table}\footnotesize
\centering
 \caption{
 \label{table-mf}
 \textbf{Quantitative reconstruction results on \textit{Multiface}.} NFR outperforms COMA, Neural3DMM, and SpiralNet++ on all metrics by a large margin. See Fig.~\ref{Fig:MF_colored} and Fig.~\ref{Fig:MF_detailed} for a qualitative comparison.  }
\begin{tabular}{ l  c  c  c  } 
  \toprule
Method & Mean (\textit{mm})   & Median  (\textit{mm}) & 90\%  (\textit{mm})\\
 \midrule
 COMA & 2.324 $\pm$ 1.724 & 1.858 & 4.380\\
 Neural3DMM & 1.254 $\pm$ 1.092 & 0.951 & 2.406\\
 SpiralNet++ & 1.256 $\pm$ 1.105 & 0.944  & 2.438\\
 NFR (\emph{Multiface}) & 1.005 $\pm$ 0.824 & 0.772 & 1.895 \\
 NFR &  \textbf{0.879 $\pm$ 0.727} & \textbf{0.678} & \textbf{1.651}\\

 \bottomrule
\end{tabular}
 \end{table}

\subsection{Triangulation Invariance}
\label{subsec:tri-agnostic}
 Our model is composed of multiple triangulation-agnostic components: The $CNN$ takes fixed-size rendered images as inputs; $DN_e$ and $DN_i$ are agnostic to the mesh templates \cite{sharp2022diffusionnet}; the MLP has shared parameters across all the input triangles. Thus, our model naturally supports meshes with different resolutions and connectivity. To quantitatively evaluate this property, we apply our model on \emph{ICT-Real-AU} with a high-resolution mesh template and different identities. The template has 10,089 vertices and 19,758 triangles. Table.~\ref{tab:triangle-agnostic} quantitatively validate that NFR has minimal performance degradation on different triangulations, despite the fact that it is only trained on the original mesh template with less than half of vertices and triangles. Sec.~\ref{subsec:retargeting} shows qualitatively that our model can transfer expression to in-the-wild meshes, even raw scans with a significant amount of noise.

\begin{table}\footnotesize
\centering
 \caption{
 \label{tab:triangle-agnostic}
 \textbf{Triangulation Invariance.} We compare the inverse-rigging performance of NFR on the trained mesh template (Original) and a retriangulated version (High-resolution) of \emph{ICT-Real-AU}. NFR performs equally well on both datasets, even though it is not trained with the retriangulated template. }
\begin{tabular}{ llcccc } 
 \toprule
 Method & Templates & Mean (\textit{mm})   & Median  (\textit{mm}) & 90\%  (\textit{mm})\\
 \midrule  \multirow{2}{*}{Ours (ICT rig)} &
 Original & 0.378 $\pm$ 0.467 & 0.225 & 0.860\\ 
 &
 High-resolution & 0.343 $\pm$ 0.416 & 0.212 & 0.774\\ 
 \hline
\multirow{2}{*}{Ours (NFR)}  & Original & 0.443 $\pm$ 0.454 & 0.307 & 0.920\\
 & High-resolution & 0.444 $\pm$ 0.436 & 0.312 & 0.930\\
 \bottomrule
\end{tabular}
 \end{table}

\subsection{Retargeting in the Wild}
\label{subsec:retargeting}
Given the target mesh of an arbitrary subject with unknown expressions, \emph{retargeting} is a task to interpret the target facial expression and transfer it to another subject. Note that this task is more challenging than a typical deformation transfer task where the neutral target mesh is known.

\begin{figure*}[t]
  \centering
    \includegraphics[width=0.75\linewidth]{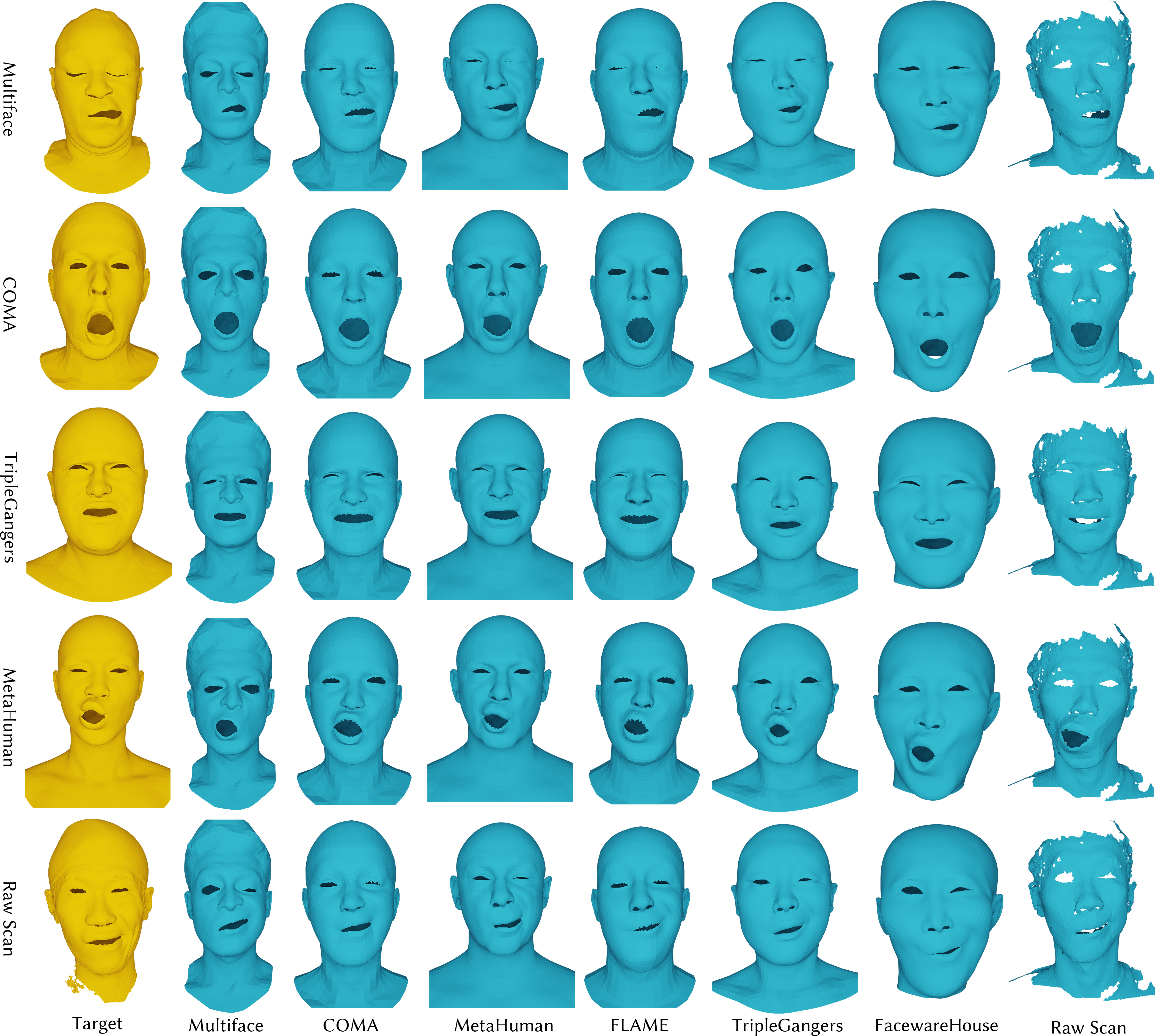}
  \caption{
  \label{Fig:in-the-wild}
  \textbf{In-the-wild retargeting}. 
\textcolor{myyellow}{Yellow}: target expression of  non-rigged face meshes with complex expressions composed of unknown AUs. 
  \textcolor{cyan}{Cyan}: retargeted results. NFR is robust to variations in identity, expression, triangulation, and mesh quality.}
\end{figure*}

In Fig.~\ref{Fig:in-the-wild}, we test the generalizability of our model's retargeting with various identities, expressions, triangulations, and mesh quality from different face models and datasets. The retargeted results to various neutral face meshes successfully capture our high-level perception of the target expressions with natural deformations, though the ground-truth AUs of the target expressions are unknown. We retarget an ARKit face tracking sequence and a \emph{Multiface} speaking sequence to these in-the-wild meshes. Please refer to the supplementary video for details. 

\subsection{Human-Friendly Editing}
\label{subsec:editing}

\begin{figure*}[t]
  \centering
    \includegraphics[width=\textwidth]{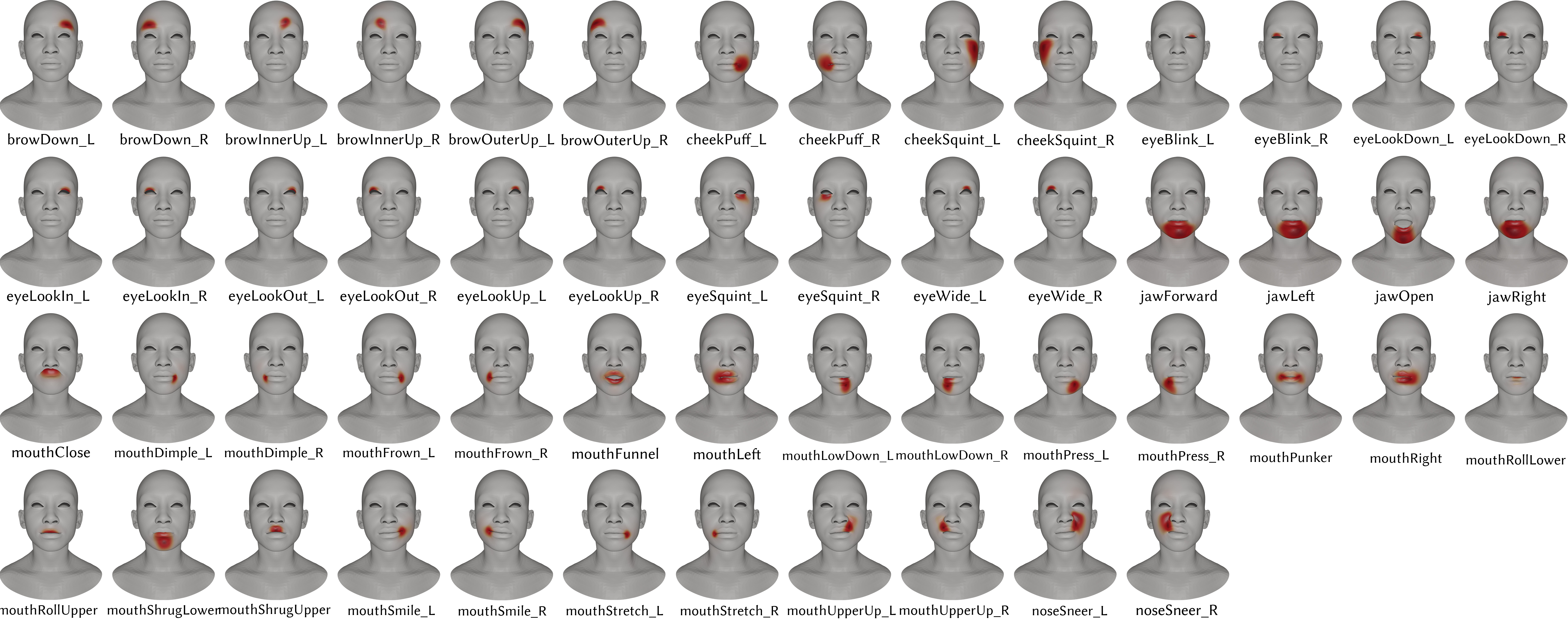}
  \caption{\textbf{Disentangled latent code visualization.} First 53 activations of our latent codes on a \emph{MetaHuman} mesh.}
  \label{fig:blendshape}
 \end{figure*}

NFR behaves as a user-friendly rig to edit the expressions with the latent code activation intensities.  Fig.~\ref{Fig:editing} shows the series of such controllable manual edits, starting off from the latent code inferred by the inverse rigging, then increasing/decreasing the latent activations corresponding to FACS AUs. We show all the 53 FACS-like control codes of our latent space in Fig.~\ref{fig:blendshape}. Note that each activation follows the semantics of the pre-defined FACS.

\subsection{Non-Linear Deformations}
\label{subsec:non-linear}
\begin{figure}[H]
  \centering
    \includegraphics[width=0.9\linewidth]{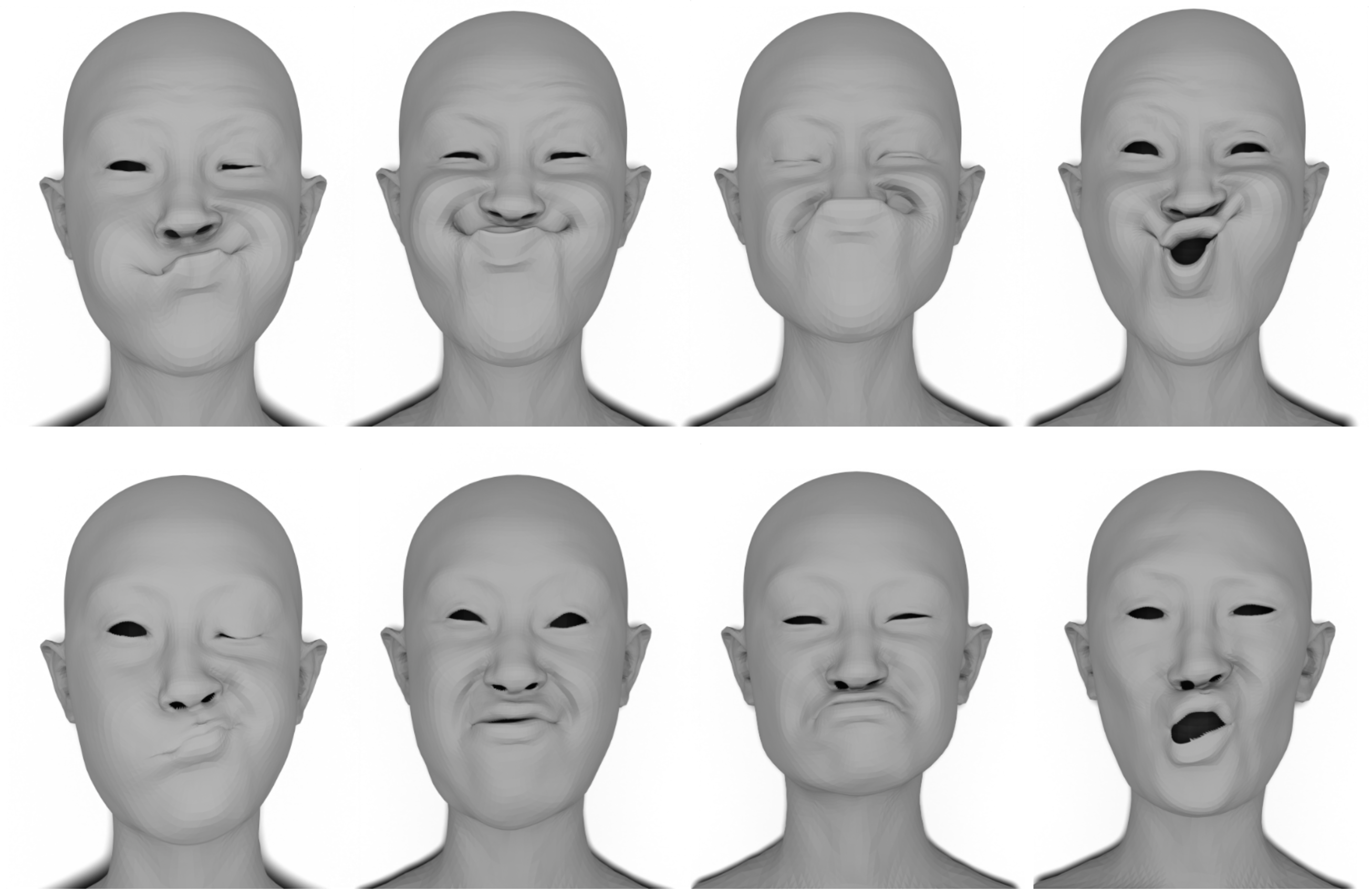}
  \caption{\textbf{Linear (top) vs. non-linear (bottom) models.} Simultaneous AU activation in linear models such as \emph{ICT FaceKit} ends up with unnatural deformations. The non-linearity in NFR helps ensure rational deformations.}
  \label{Fig:reasonable}
\end{figure}

Along with the interpretable latent code learned from \emph{ICT FaceKit}, our model also captures natural, non-linear deformations from \emph{Multiface}. We show this in Fig.~\ref{Fig:reasonable} where many AUs are activated simultaneously. Linear 3DMMs struggle with such complex activations as they simply add more offsets on top of offsets, resulting in implausible deformations. A usual workaround  is to activate manually-sculpted corrective shapes  of activation combinations to introduce non-linearity~\cite{lewis2014practice}. Our model maintains natural, non-linear deformations without such ad-hoc sculpting.

\subsection{Ablations}
\label{subsec:Ablations}
We perform a series of ablation studies to demonstrate the effectiveness of our key designs. 

\begin{figure}[H]
  \centering
    \includegraphics[width=0.9\linewidth]{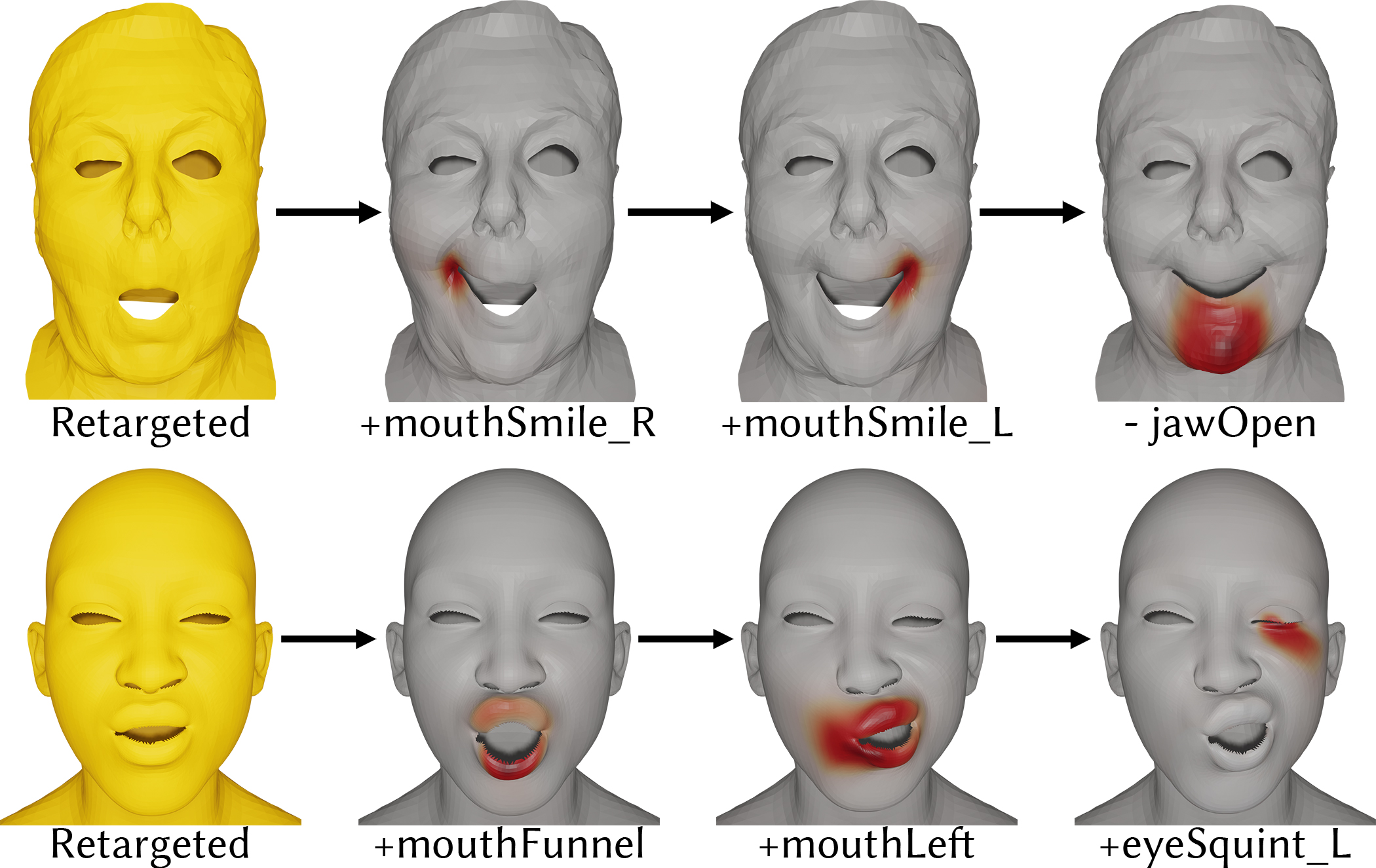}
  \caption{
  \label{Fig:editing}
  \textbf{Human-friendly editing}. \textcolor{myyellow}{Yellow}:  non-rigged face mesh. \textcolor{gray}{Gray}: Meshes after editing AUs.  Vertices are colored by the Euclidean distance before and after each modification.}
\end{figure}

\begin{figure}[H]
\centering
\includegraphics[width=0.9\linewidth]{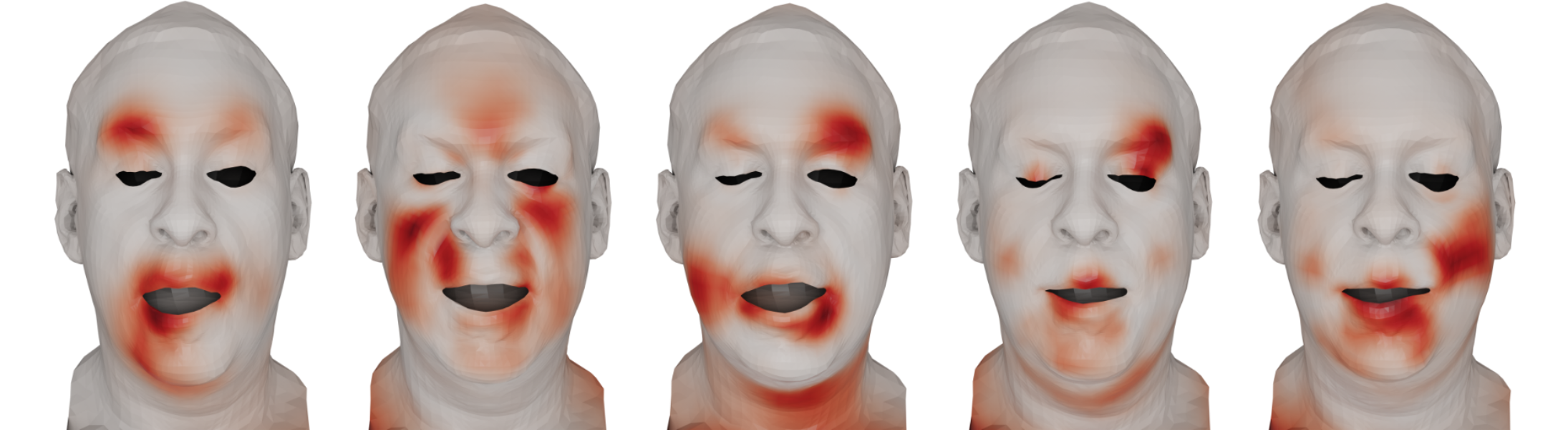}
\caption{\textbf{Ablation showing the importance of synthetic data (\emph{ICT FaceKit}).} We show the effects of the first five AUs on \emph{Multiface} without training on \emph{ICT FaceKit}. The deformations are highly entangled and impractical for human editing.}
\label{Fig:ablation_ICT}
\end{figure}

\paragraph{Multi-dataset Training Scheme}  The training scheme is the core of our method for realistic expression generation and interpretable editing. Fig.~\ref{Fig:editing} demonstrates the effectiveness of our user-friendly editing. In contrast, in Fig.~\ref{Fig:ablation_ICT} we show the activation of the first five individual AUs, without training on the ICT datasets.  The deformed vertices are highly entangled.  Without the direct supervision of ICT AUs, making adjustments by tweaking the expression codes is impractical.
We then show quantitatively in  Table.~\ref{tab:ab-multiface} that our model has poor quality on the reconstructed expressions without training on the real world \emph{Multiface} data. 

\begin{table}\footnotesize
\centering
 \caption{
 \label{tab:ab-multiface}
 \textbf{Ablation showing the importance of real data (\emph{Multiface})}. We evaluate on \emph{Multiface} test set with and without training on \emph{Multiface} real scans.}
\begin{tabular}{ lcccc } 
 \toprule
 Method & Mean (\textit{mm})   & Median  (\textit{mm}) & 90\%  (\textit{mm})\\
 
 \midrule
   With \emph{Multiface} &  \textbf{0.879 $\pm$ 0.727} & \textbf{0.678} & \textbf{1.651}\\
   Without \emph{Multiface} & 1.470 $\pm$ 1.319 & 1.088 & 2.860\\

 \bottomrule
\end{tabular}
 \end{table} 
 
\begin{figure}[H]
\centering
\includegraphics[width=0.9\linewidth]{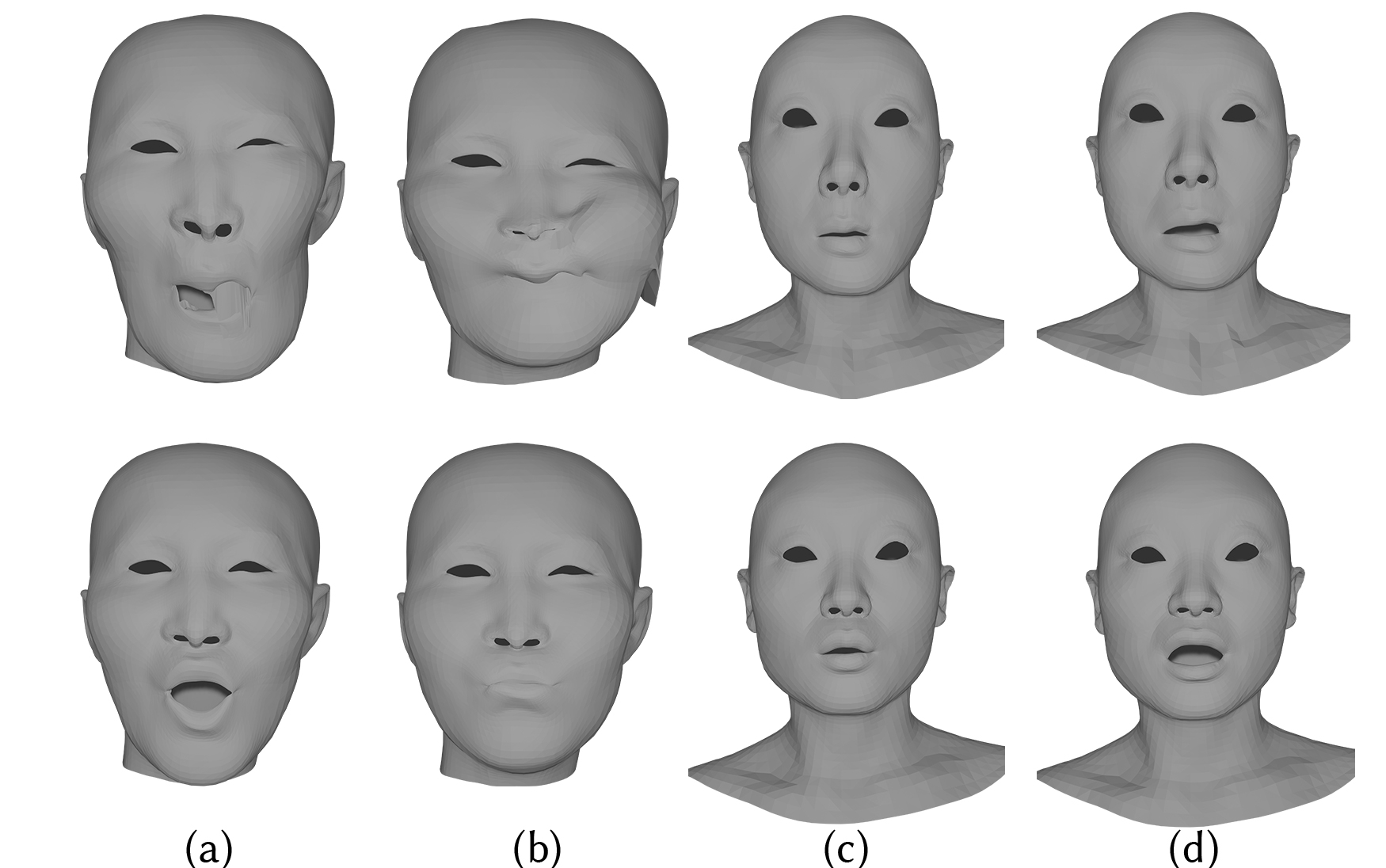}
\caption{\textbf{Qualitative ablation of our data augmentation.} Without (\textbf{top}) vs. with (\textbf{bottom}) data augmentation. (a), (b) from \emph{FaceWarehouse} , (c), (d) from \emph{Triplegangers}. Our data augmentation is critical to NFR working on these in-the-wild meshes.}
\label{fig:argumentation-ab}
\end{figure}

\paragraph{Data Augmentation} We show in Fig.~\ref{fig:argumentation-ab} NFR with/without data augmentation to retarget meshes from two different datasets: \emph{FaceWarehouse}~\cite{cao2013facewarehouse} and \emph{Triplegangers}. Without training on the augmented datasets in Sec~\ref{subsec:data-augmentation}, the model fails to apply the correct deformations to these in-the-wild meshes. 

\begin{figure}[H]
\centering
        \begin{subfigure}[b]{0.45\linewidth}
                \includegraphics[width=\linewidth]{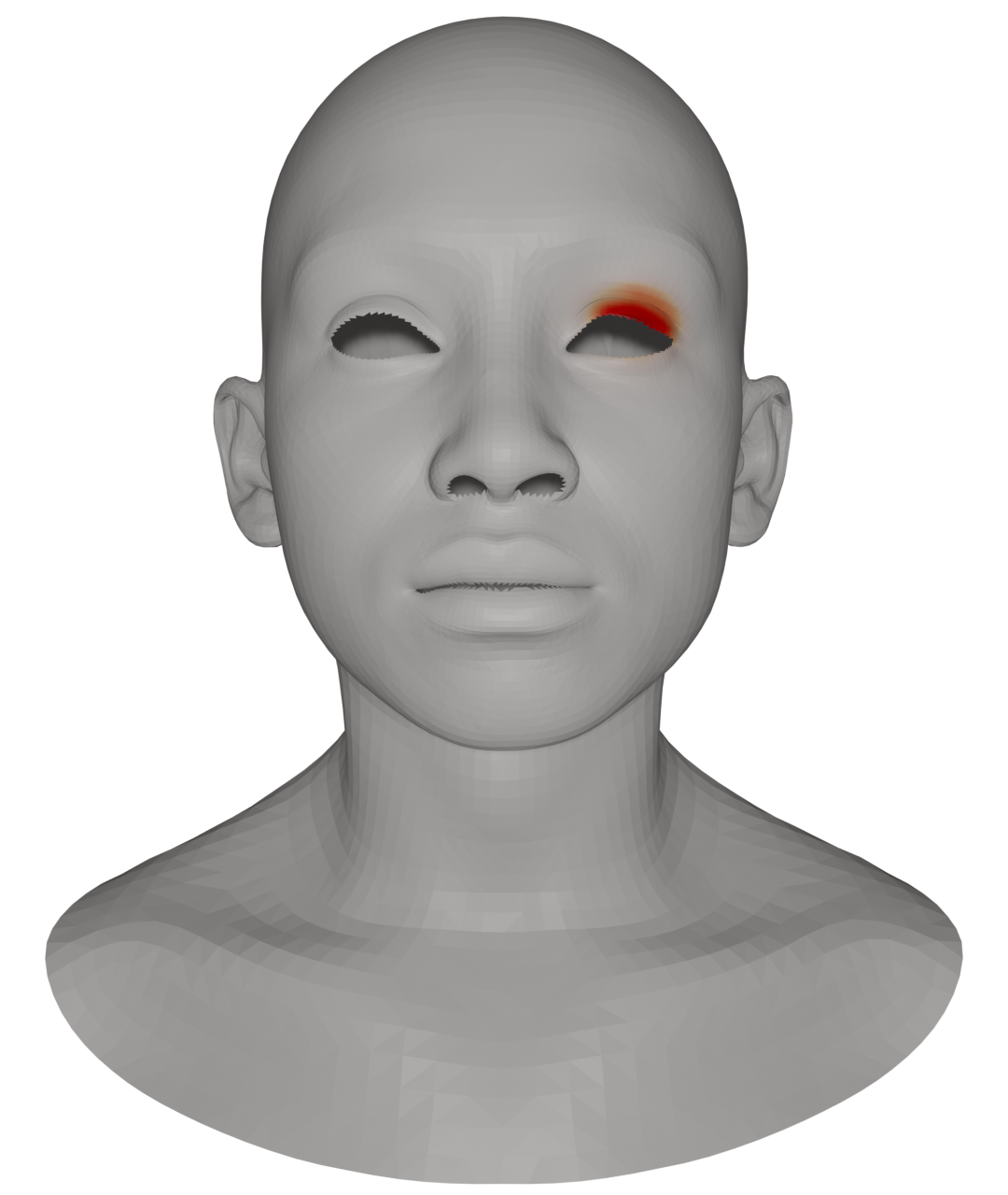}
                \caption{Expression \textit{eyeLookIn\_L} lowers the left eye lid.}

        \end{subfigure}\hspace{0.2cm}%
        \begin{subfigure}[b]{0.45\linewidth}
                \includegraphics[width=\linewidth]{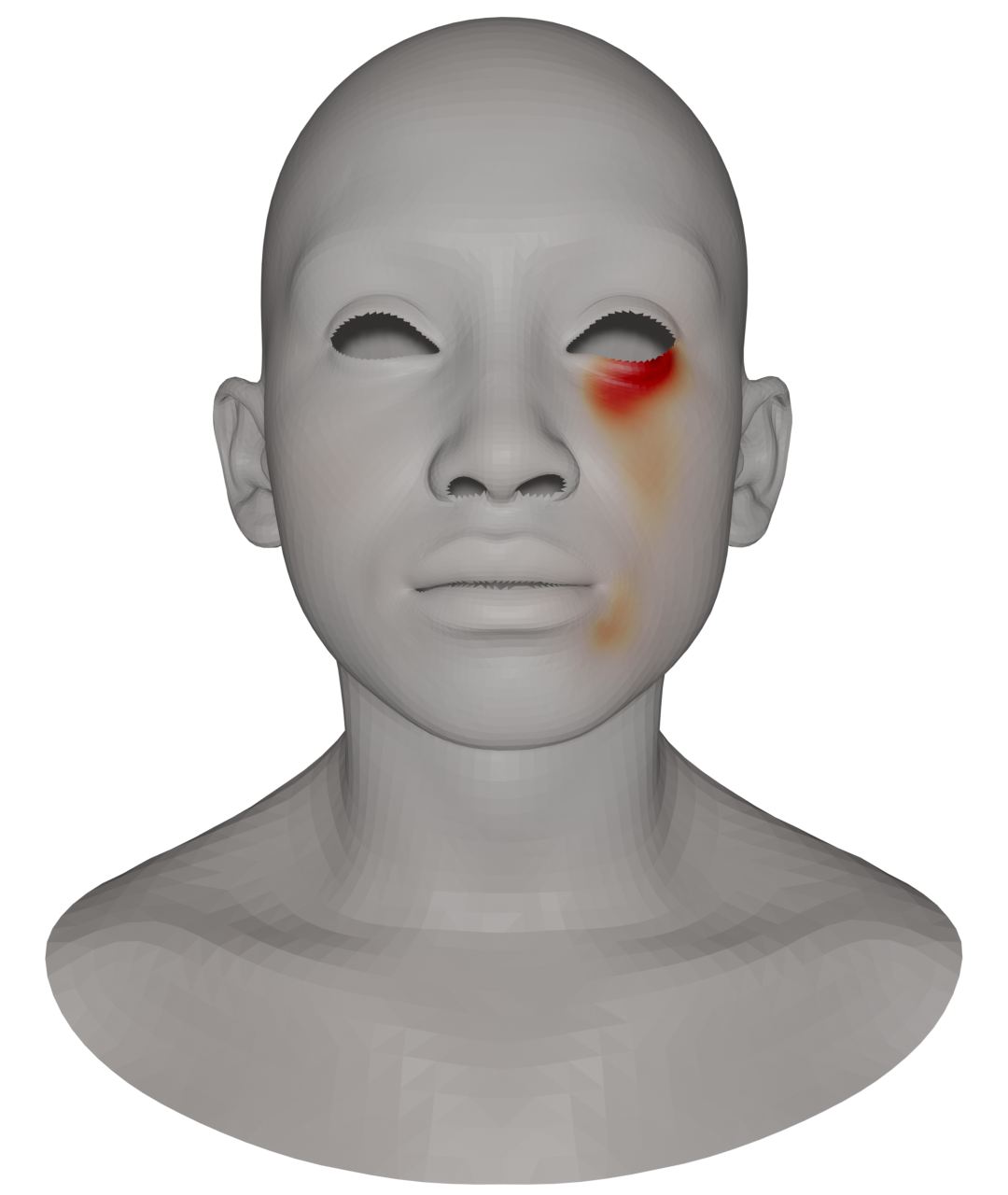}
                \caption{The same deformation heatmap without $z_{ext}$.}

        \end{subfigure}%
        \\
        \begin{subfigure}[b]{0.45\linewidth}
                \includegraphics[width=\linewidth]{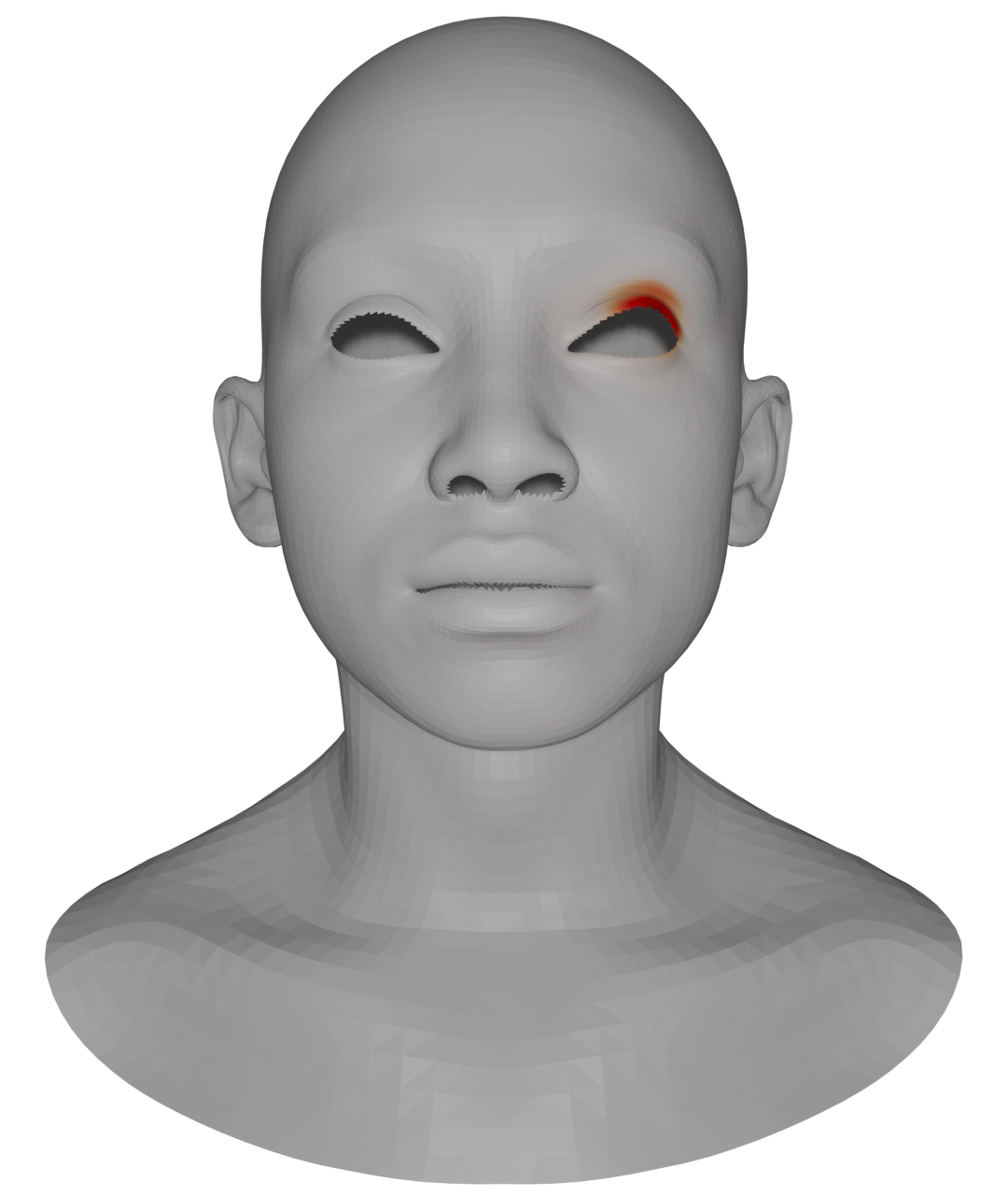}
                \caption{Expression \textit{eyeLookUp\_L} raises the left eye lid.}

        \end{subfigure}\hspace{0.2cm}%
        \begin{subfigure}[b]{0.45\linewidth}
                \includegraphics[width=\linewidth]{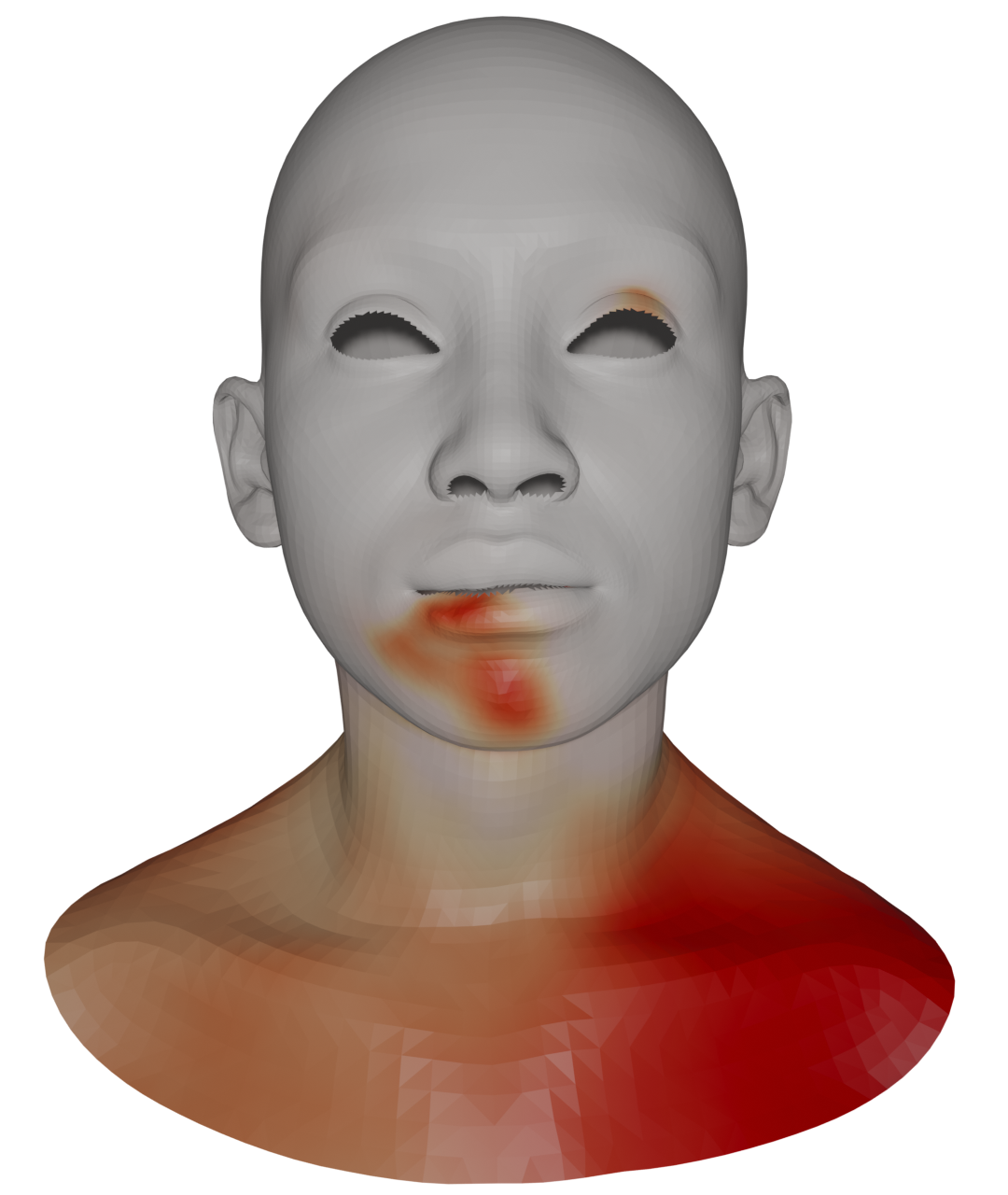}
                \caption{The same deformation heatmap without $z_{ext}$.}

        \end{subfigure}
\caption{\label{fig:z-extend-ab}\textbf{Effects of $z_{ext}$.} Here we plot the deformation heatmaps of two expression codes. \textit{\textbf{Left:}} with the extended latent space $z_{ext}$, $z_{FACS}$ successfully captures the disentangled deformation with correct semantics. \textit{\textbf{Right:}} Without $z_{ext}, z_{FACS}$ gets contaminated with entangled deformation in \emph{Multiface}.}

\end{figure}

\paragraph{Extended latent space} We expand the latent expression space to be 128-dimensional, where the first 53-dimensional vector, denoted as $z_{FACS}$ is supervised by the ICT expression codes. The remaining $z_{ext}$ is learned from data. Fig.~\ref{fig:z-extend-ab} compares individual activations of $z_{FACS}$ between NFR and a similar model without $z_{ext}$. Both models supervise $z_{FACS}$ to follow the ICT rig.  Since \emph{Multiface} contains deformation patterns that are not covered by the ICT rig space, the model without $z_{ext}$ adapts some AUs, \textit{e.g.} `eyeLookUp\_L', to capture these new patterns (\textit{e.g.} neck deformation). With the additional expression space of $z_{ext}$, NFR can capture the \emph{Multiface} deformations and maintain maximum interpretability. 

\paragraph{Network Structures} The $CNN$ serves to generalize NFR to the translation and scale introduced in data augmentation for better in-the-wild applicability. $\text{DN}_e$ and $\text{DN}_i$ are crucial for solving the correct expression codes and transferring them to in-the-wild meshes. Table.~\ref{tab:ab-dn} compares a model without the CNN features and another one that replaces the two DNs by PointNets (PN) \cite{qi2017pointnet}. The performance drops significantly. 

\begin{table}\footnotesize
\centering
 \caption{
 \label{tab:ab-dn}
 \textbf{Ablation study on the network structures}. Without the CNN features, the model struggles to deal with the random shift and scale introduced in data augmentation. When replacing the DN layers with PN, the model fails to solve for the correct expression codes.}
\begin{tabular}{ llcccc } 
 \toprule
Dataset & Method & Mean (\textit{mm})   & Median  (\textit{mm}) & 90\%  (\textit{mm})\\
 
 \midrule
 \multirow{3}{*}{ICT-Real-AU} 
 & Ours (w/o CNN) & 1.034 $\pm$ 0.985 & 0.748 & 2.053\\ 
    &  Ours (PN) & 0.710 $\pm$ 0.756 & 0.483 & 1.502\\
    &  Ours (DN)  & \textbf{0.443} $\pm$ \textbf{0.454} & \textbf{0.307} & \textbf{0.920}\\
\hline
  \multirow{3}{*}{Multiface} 
    & Ours (w/o CNN) & 1.559 $\pm$ 1.278 & 1.184 & 3.011\\ 
    & Ours (PN) &  1.408 $\pm$ 1.207 & 1.050 & 2.758\\
    & Ours (DN) &  \textbf{0.879} $\pm$ \textbf{0.727} & \textbf{0.678} & \textbf{1.651}\\

 \bottomrule
\end{tabular}
 \end{table}  

\section{Conclusions}
    We have presented \emph{Neural Face Rigging}, a novel learning approach to instantly rig and retarget 3D facial meshes in the wild with any reasonable shape variations and triangulation. The key technical contribution is the multi-stage training combining the advantages of different face datasets to learn interpretable and editable latent code over high-fidelity facial deformations. While our model is unique in its generalization to mesh triangulation, it learns better facial deformation than other methods that require fixed triangulation.

\paragraph{Limitations and Future Work}
Although our model provides a triangulation-agnostic facial rigging and retargeting pipeline, users still need a standardization step  by removing internal structures around the eyes and mouth. Segmenting the facial regions could automate the process.

We chose to learn an ARKit-compatible FACS rig for its popularity and accessibility. In theory, NFR could learn any other arbitrary rig parameterizations such as \textit{MetaHuman}, though this is not tested because of the limited access to such data.

With an interpretable, controllable latent space and triangulation invariance, our model can serve as a backbone for various facial animation tasks such as talking face generation. Including the appearance model for photo-realistic expression generation is also worth exploring. 

Human faces are sensitive subjects. We must take precautions deploying our model, with in-depth studies on the bias to different human attributes, e.g. age, gender, and ethnicity. We hope this work can contribute to the community by minimizing such bias with its ability to combine multiple face datasets and perform instant rigging on in-the-wild meshes.

    \begin{acks}
This research is supported by 
Innovation and Technology Commission
(Ref:ITS/319/21FP) and Research Grant Council (Ref:  
17210222), Hong Kong.   
\end{acks}

\bibliographystyle{ACM-Reference-Format}
\bibliography{ref}


\begin{thebibliography}{44}


\ifx \showCODEN    \undefined \def \showCODEN     #1{\unskip}     \fi
\ifx \showDOI      \undefined \def \showDOI       #1{#1}\fi
\ifx \showISBNx    \undefined \def \showISBNx     #1{\unskip}     \fi
\ifx \showISBNxiii \undefined \def \showISBNxiii  #1{\unskip}     \fi
\ifx \showISSN     \undefined \def \showISSN      #1{\unskip}     \fi
\ifx \showLCCN     \undefined \def \showLCCN      #1{\unskip}     \fi
\ifx \shownote     \undefined \def \shownote      #1{#1}          \fi
\ifx \showarticletitle \undefined \def \showarticletitle #1{#1}   \fi
\ifx \showURL      \undefined \def \showURL       {\relax}        \fi
\providecommand\bibfield[2]{#2}
\providecommand\bibinfo[2]{#2}
\providecommand\natexlab[1]{#1}
\providecommand\showeprint[2][]{arXiv:#2}

\bibitem[Aigerman et~al\mbox{.}(2022)]%
        {10.1145/3528223.3530141}
\bibfield{author}{\bibinfo{person}{Noam Aigerman}, \bibinfo{person}{Kunal
  Gupta}, \bibinfo{person}{Vladimir~G. Kim}, \bibinfo{person}{Siddhartha
  Chaudhuri}, \bibinfo{person}{Jun Saito}, {and} \bibinfo{person}{Thibault
  Groueix}.} \bibinfo{year}{2022}\natexlab{}.
\newblock \showarticletitle{Neural Jacobian Fields: Learning Intrinsic Mappings
  of Arbitrary Meshes}.
\newblock \bibinfo{journal}{\emph{ACM Trans. Graph.}} \bibinfo{volume}{41},
  \bibinfo{number}{4}, Article \bibinfo{articleno}{109} (\bibinfo{date}{jul}
  \bibinfo{year}{2022}), \bibinfo{numpages}{17}~pages.
\newblock
\showISSN{0730-0301}
\urldef\tempurl%
\url{https://doi.org/10.1145/3528223.3530141}
\showDOI{\tempurl}


\bibitem[Bailey et~al\mbox{.}(2020)]%
        {bailey2020fast}
\bibfield{author}{\bibinfo{person}{Stephen~W Bailey}, \bibinfo{person}{Dalton
  Omens}, \bibinfo{person}{Paul Dilorenzo}, {and} \bibinfo{person}{James~F
  O'Brien}.} \bibinfo{year}{2020}\natexlab{}.
\newblock \showarticletitle{Fast and deep facial deformations}.
\newblock \bibinfo{journal}{\emph{ACM Transactions on Graphics (TOG)}}
  \bibinfo{volume}{39}, \bibinfo{number}{4} (\bibinfo{year}{2020}),
  \bibinfo{pages}{94--1}.
\newblock


\bibitem[Blanz and Vetter(1999)]%
        {blanz1999morphable}
\bibfield{author}{\bibinfo{person}{Volker Blanz} {and} \bibinfo{person}{Thomas
  Vetter}.} \bibinfo{year}{1999}\natexlab{}.
\newblock \showarticletitle{A morphable model for the synthesis of 3D faces}.
  In \bibinfo{booktitle}{\emph{Proceedings of the 26th annual conference on
  Computer graphics and interactive techniques}}. \bibinfo{pages}{187--194}.
\newblock


\bibitem[Bouritsas et~al\mbox{.}(2019)]%
        {bouritsas2019neural}
\bibfield{author}{\bibinfo{person}{Giorgos Bouritsas}, \bibinfo{person}{Sergiy
  Bokhnyak}, \bibinfo{person}{Stylianos Ploumpis}, \bibinfo{person}{Michael
  Bronstein}, {and} \bibinfo{person}{Stefanos Zafeiriou}.}
  \bibinfo{year}{2019}\natexlab{}.
\newblock \showarticletitle{Neural 3d morphable models: Spiral convolutional
  networks for 3d shape representation learning and generation}. In
  \bibinfo{booktitle}{\emph{Proceedings of the IEEE/CVF International
  Conference on Computer Vision}}. \bibinfo{pages}{7213--7222}.
\newblock


\bibitem[Brunton et~al\mbox{.}(2014)]%
        {brunton2014multilinear}
\bibfield{author}{\bibinfo{person}{Alan Brunton}, \bibinfo{person}{Timo
  Bolkart}, {and} \bibinfo{person}{Stefanie Wuhrer}.}
  \bibinfo{year}{2014}\natexlab{}.
\newblock \showarticletitle{Multilinear wavelets: A statistical shape space for
  human faces}. In \bibinfo{booktitle}{\emph{European Conference on Computer
  Vision}}. Springer, \bibinfo{pages}{297--312}.
\newblock


\bibitem[Cao et~al\mbox{.}(2013)]%
        {cao2013facewarehouse}
\bibfield{author}{\bibinfo{person}{Chen Cao}, \bibinfo{person}{Yanlin Weng},
  \bibinfo{person}{Shun Zhou}, \bibinfo{person}{Yiying Tong}, {and}
  \bibinfo{person}{Kun Zhou}.} \bibinfo{year}{2013}\natexlab{}.
\newblock \showarticletitle{Facewarehouse: A 3d facial expression database for
  visual computing}.
\newblock \bibinfo{journal}{\emph{IEEE Transactions on Visualization and
  Computer Graphics}} \bibinfo{volume}{20}, \bibinfo{number}{3}
  (\bibinfo{year}{2013}), \bibinfo{pages}{413--425}.
\newblock


\bibitem[Cetinaslan and Orvalho(2020a)]%
        {cetinaslan2020sketching}
\bibfield{author}{\bibinfo{person}{Ozan Cetinaslan} {and}
  \bibinfo{person}{Ver{\'o}nica Orvalho}.} \bibinfo{year}{2020}\natexlab{a}.
\newblock \showarticletitle{Sketching Manipulators for Localized Blendshape
  Editing}.
\newblock \bibinfo{journal}{\emph{Graphical Models}}  \bibinfo{volume}{108}
  (\bibinfo{year}{2020}), \bibinfo{pages}{101059}.
\newblock


\bibitem[Cetinaslan and Orvalho(2020b)]%
        {cetinaslan2020stabilized}
\bibfield{author}{\bibinfo{person}{Ozan Cetinaslan} {and}
  \bibinfo{person}{Ver{\'o}nica Orvalho}.} \bibinfo{year}{2020}\natexlab{b}.
\newblock \showarticletitle{Stabilized blendshape editing using localized
  Jacobian transpose descent}.
\newblock \bibinfo{journal}{\emph{Graphical Models}}  \bibinfo{volume}{112}
  (\bibinfo{year}{2020}), \bibinfo{pages}{101091}.
\newblock


\bibitem[Chandran et~al\mbox{.}(2020)]%
        {chandran2020semantic}
\bibfield{author}{\bibinfo{person}{Prashanth Chandran}, \bibinfo{person}{Derek
  Bradley}, \bibinfo{person}{Markus Gross}, {and} \bibinfo{person}{Thabo
  Beeler}.} \bibinfo{year}{2020}\natexlab{}.
\newblock \showarticletitle{Semantic deep face models}. In
  \bibinfo{booktitle}{\emph{2020 International Conference on 3D Vision (3DV)}}.
  IEEE, \bibinfo{pages}{345--354}.
\newblock


\bibitem[Chandran et~al\mbox{.}(2022a)]%
        {chandran2022facialanimation}
\bibfield{author}{\bibinfo{person}{Prashanth Chandran},
  \bibinfo{person}{Gaspard Zoss}, \bibinfo{person}{Markus Gross},
  \bibinfo{person}{Paulo Gotardo}, {and} \bibinfo{person}{Derek Bradley}.}
  \bibinfo{year}{2022}\natexlab{a}.
\newblock \showarticletitle{Facial Animation with Disentangled Identity and
  Motion using Transformers}.
\newblock \bibinfo{journal}{\emph{ACM/Eurographics Symposium on Computer
  Animation}} (\bibinfo{year}{2022}).
\newblock


\bibitem[Chandran et~al\mbox{.}(2022b)]%
        {chandran2022shape}
\bibfield{author}{\bibinfo{person}{Prashanth Chandran},
  \bibinfo{person}{Gaspard Zoss}, \bibinfo{person}{Markus Gross},
  \bibinfo{person}{Paulo Gotardo}, {and} \bibinfo{person}{Derek Bradley}.}
  \bibinfo{year}{2022}\natexlab{b}.
\newblock \showarticletitle{Shape Transformers: Topology-Independent 3D Shape
  Models Using Transformers}. In \bibinfo{booktitle}{\emph{Computer Graphics
  Forum}}, Vol.~\bibinfo{volume}{41}. Wiley Online Library,
  \bibinfo{pages}{195--207}.
\newblock


\bibitem[Choe and Ko(2006)]%
        {choe2006analysis}
\bibfield{author}{\bibinfo{person}{Byoungwon Choe} {and}
  \bibinfo{person}{Hyeong-Seok Ko}.} \bibinfo{year}{2006}\natexlab{}.
\newblock \showarticletitle{Analysis and synthesis of facial expressions with
  hand-generated muscle actuation basis}.
\newblock In \bibinfo{booktitle}{\emph{ACM SIGGRAPH 2006 Courses}}.
  \bibinfo{pages}{21--es}.
\newblock


\bibitem[Choe et~al\mbox{.}(2001)]%
        {choe2001performance}
\bibfield{author}{\bibinfo{person}{Byoungwon Choe}, \bibinfo{person}{Hanook
  Lee}, {and} \bibinfo{person}{Hyeong-Seok Ko}.}
  \bibinfo{year}{2001}\natexlab{}.
\newblock \showarticletitle{Performance-driven muscle-based facial animation}.
\newblock \bibinfo{journal}{\emph{The Journal of Visualization and Computer
  Animation}} \bibinfo{volume}{12}, \bibinfo{number}{2} (\bibinfo{year}{2001}),
  \bibinfo{pages}{67--79}.
\newblock


\bibitem[Choi et~al\mbox{.}(2022)]%
        {choi2022animatomy}
\bibfield{author}{\bibinfo{person}{Byungkuk Choi}, \bibinfo{person}{Haekwang
  Eom}, \bibinfo{person}{Benjamin Mouscadet}, \bibinfo{person}{Stephen
  Cullingford}, \bibinfo{person}{Kurt Ma}, \bibinfo{person}{Stefanie Gassel},
  \bibinfo{person}{Suzi Kim}, \bibinfo{person}{Andrew Moffat},
  \bibinfo{person}{Millicent Maier}, \bibinfo{person}{Marco Revelant},
  \bibinfo{person}{Joe Letteri}, {and} \bibinfo{person}{Karan Singh}.}
  \bibinfo{year}{2022}\natexlab{}.
\newblock \showarticletitle{Animatomy: An Animator-Centric, Anatomically
  Inspired System for 3D Facial Modeling, Animation and Transfer}. In
  \bibinfo{booktitle}{\emph{SIGGRAPH Asia 2022 Conference Papers}} (Daegu,
  Republic of Korea) \emph{(\bibinfo{series}{SA '22})}.
  \bibinfo{publisher}{Association for Computing Machinery},
  \bibinfo{address}{New York, NY, USA}, Article \bibinfo{articleno}{16},
  \bibinfo{numpages}{9}~pages.
\newblock
\showISBNx{9781450394703}
\urldef\tempurl%
\url{https://doi.org/10.1145/3550469.3555398}
\showDOI{\tempurl}


\bibitem[Egger et~al\mbox{.}(2020)]%
        {3DMM_survey}
\bibfield{author}{\bibinfo{person}{Bernhard Egger}, \bibinfo{person}{William
  A.~P. Smith}, \bibinfo{person}{Ayush Tewari}, \bibinfo{person}{Stefanie
  Wuhrer}, \bibinfo{person}{Michael Zollhoefer}, \bibinfo{person}{Thabo
  Beeler}, \bibinfo{person}{Florian Bernard}, \bibinfo{person}{Timo Bolkart},
  \bibinfo{person}{Adam Kortylewski}, \bibinfo{person}{Sami Romdhani},
  \bibinfo{person}{Christian Theobalt}, \bibinfo{person}{Volker Blanz}, {and}
  \bibinfo{person}{Thomas Vetter}.} \bibinfo{year}{2020}\natexlab{}.
\newblock \showarticletitle{3D Morphable Face Models - Past, Present and
  Future}.
\newblock \bibinfo{journal}{\emph{ACM Transactions on Graphics}}
  \bibinfo{volume}{39}, \bibinfo{number}{5} (\bibinfo{date}{August}
  \bibinfo{year}{2020}).
\newblock
\urldef\tempurl%
\url{https://doi.org/10.1145/3395208}
\showDOI{\tempurl}


\bibitem[Ekman and Friesen(1978)]%
        {Ekman1978FacialAC}
\bibfield{author}{\bibinfo{person}{Paul Ekman} {and}
  \bibinfo{person}{Wallace~V. Friesen}.} \bibinfo{year}{1978}\natexlab{}.
\newblock \showarticletitle{Facial action coding system: a technique for the
  measurement of facial movement}. In \bibinfo{booktitle}{\emph{Consulting
  Psychologists Press}}.
\newblock


\bibitem[Gao et~al\mbox{.}(2018)]%
        {gao2018automatic}
\bibfield{author}{\bibinfo{person}{Lin Gao}, \bibinfo{person}{Jie Yang},
  \bibinfo{person}{Yi-Ling Qiao}, \bibinfo{person}{Yu-Kun Lai},
  \bibinfo{person}{Paul~L Rosin}, \bibinfo{person}{Weiwei Xu}, {and}
  \bibinfo{person}{Shihong Xia}.} \bibinfo{year}{2018}\natexlab{}.
\newblock \showarticletitle{Automatic unpaired shape deformation transfer}.
\newblock \bibinfo{journal}{\emph{ACM Transactions on Graphics (TOG)}}
  \bibinfo{volume}{37}, \bibinfo{number}{6} (\bibinfo{year}{2018}),
  \bibinfo{pages}{1--15}.
\newblock


\bibitem[Gong et~al\mbox{.}(2019)]%
        {gong2019spiralnet++}
\bibfield{author}{\bibinfo{person}{Shunwang Gong}, \bibinfo{person}{Lei Chen},
  \bibinfo{person}{Michael Bronstein}, {and} \bibinfo{person}{Stefanos
  Zafeiriou}.} \bibinfo{year}{2019}\natexlab{}.
\newblock \showarticletitle{Spiralnet++: A fast and highly efficient mesh
  convolution operator}. In \bibinfo{booktitle}{\emph{Proceedings of the
  IEEE/CVF International Conference on Computer Vision Workshops}}.
  \bibinfo{pages}{0--0}.
\newblock


\bibitem[Huber et~al\mbox{.}(2016)]%
        {HuberP2016AM3M}
\bibfield{author}{\bibinfo{person}{P Huber}, \bibinfo{person}{G Hu},
  \bibinfo{person}{R Tena}, \bibinfo{person}{P Mortazavian}, \bibinfo{person}{P
  Koppen}, \bibinfo{person}{WJ Christmas}, \bibinfo{person}{M Ratsch}, {and}
  \bibinfo{person}{J Kittler}.} \bibinfo{year}{2016}\natexlab{}.
\newblock \bibinfo{title}{A Multiresolution 3D Morphable Face Model and Fitting
  Framework}.
\newblock
\newblock


\bibitem[Jiang et~al\mbox{.}(2019)]%
        {jiang2019disentangled}
\bibfield{author}{\bibinfo{person}{Zi-Hang Jiang}, \bibinfo{person}{Qianyi Wu},
  \bibinfo{person}{Keyu Chen}, {and} \bibinfo{person}{Juyong Zhang}.}
  \bibinfo{year}{2019}\natexlab{}.
\newblock \showarticletitle{Disentangled representation learning for 3d face
  shape}. In \bibinfo{booktitle}{\emph{Proceedings of the IEEE/CVF Conference
  on Computer Vision and Pattern Recognition}}. \bibinfo{pages}{11957--11966}.
\newblock


\bibitem[Lewis et~al\mbox{.}(2014)]%
        {lewis2014practice}
\bibfield{author}{\bibinfo{person}{John~P Lewis}, \bibinfo{person}{Ken Anjyo},
  \bibinfo{person}{Taehyun Rhee}, \bibinfo{person}{Mengjie Zhang},
  \bibinfo{person}{Frederic~H Pighin}, {and} \bibinfo{person}{Zhigang Deng}.}
  \bibinfo{year}{2014}\natexlab{}.
\newblock \showarticletitle{Practice and theory of blendshape facial models.}
\newblock \bibinfo{journal}{\emph{Eurographics (State of the Art Reports)}}
  \bibinfo{volume}{1}, \bibinfo{number}{8} (\bibinfo{year}{2014}),
  \bibinfo{pages}{2}.
\newblock


\bibitem[Lewis and Anjyo(2010)]%
        {lewis2010direct}
\bibfield{author}{\bibinfo{person}{John~P Lewis} {and}
  \bibinfo{person}{Ken-ichi Anjyo}.} \bibinfo{year}{2010}\natexlab{}.
\newblock \showarticletitle{Direct manipulation blendshapes}.
\newblock \bibinfo{journal}{\emph{IEEE Computer Graphics and Applications}}
  \bibinfo{volume}{30}, \bibinfo{number}{4} (\bibinfo{year}{2010}),
  \bibinfo{pages}{42--50}.
\newblock


\bibitem[Li et~al\mbox{.}(2010)]%
        {li2010example}
\bibfield{author}{\bibinfo{person}{Hao Li}, \bibinfo{person}{Thibaut Weise},
  {and} \bibinfo{person}{Mark Pauly}.} \bibinfo{year}{2010}\natexlab{}.
\newblock \showarticletitle{Example-based facial rigging}.
\newblock \bibinfo{journal}{\emph{Acm transactions on graphics (tog)}}
  \bibinfo{volume}{29}, \bibinfo{number}{4} (\bibinfo{year}{2010}),
  \bibinfo{pages}{1--6}.
\newblock


\bibitem[Li et~al\mbox{.}(2020)]%
        {li2020learning}
\bibfield{author}{\bibinfo{person}{Ruilong Li}, \bibinfo{person}{Karl Bladin},
  \bibinfo{person}{Yajie Zhao}, \bibinfo{person}{Chinmay Chinara},
  \bibinfo{person}{Owen Ingraham}, \bibinfo{person}{Pengda Xiang},
  \bibinfo{person}{Xinglei Ren}, \bibinfo{person}{Pratusha Prasad},
  \bibinfo{person}{Bipin Kishore}, \bibinfo{person}{Jun Xing}, {et~al\mbox{.}}}
  \bibinfo{year}{2020}\natexlab{}.
\newblock \showarticletitle{Learning formation of physically-based face
  attributes}. In \bibinfo{booktitle}{\emph{Proceedings of the IEEE/CVF
  conference on computer vision and pattern recognition}}.
  \bibinfo{pages}{3410--3419}.
\newblock


\bibitem[Li et~al\mbox{.}(2017)]%
        {li2017learning}
\bibfield{author}{\bibinfo{person}{Tianye Li}, \bibinfo{person}{Timo Bolkart},
  \bibinfo{person}{Michael~J Black}, \bibinfo{person}{Hao Li}, {and}
  \bibinfo{person}{Javier Romero}.} \bibinfo{year}{2017}\natexlab{}.
\newblock \showarticletitle{Learning a model of facial shape and expression
  from 4D scans.}
\newblock \bibinfo{journal}{\emph{ACM Trans. Graph.}} \bibinfo{volume}{36},
  \bibinfo{number}{6} (\bibinfo{year}{2017}), \bibinfo{pages}{194--1}.
\newblock


\bibitem[Monti et~al\mbox{.}(2017)]%
        {monti2017geometric}
\bibfield{author}{\bibinfo{person}{Federico Monti}, \bibinfo{person}{Davide
  Boscaini}, \bibinfo{person}{Jonathan Masci}, \bibinfo{person}{Emanuele
  Rodola}, \bibinfo{person}{Jan Svoboda}, {and} \bibinfo{person}{Michael~M
  Bronstein}.} \bibinfo{year}{2017}\natexlab{}.
\newblock \showarticletitle{Geometric deep learning on graphs and manifolds
  using mixture model cnns}. In \bibinfo{booktitle}{\emph{Proceedings of the
  IEEE conference on computer vision and pattern recognition}}.
  \bibinfo{pages}{5115--5124}.
\newblock


\bibitem[Moser et~al\mbox{.}(2021)]%
        {moser2021semi}
\bibfield{author}{\bibinfo{person}{Lucio Moser}, \bibinfo{person}{Chinyu
  Chien}, \bibinfo{person}{Mark Williams}, \bibinfo{person}{Jose Serra},
  \bibinfo{person}{Darren Hendler}, {and} \bibinfo{person}{Doug Roble}.}
  \bibinfo{year}{2021}\natexlab{}.
\newblock \showarticletitle{Semi-supervised video-driven facial animation
  transfer for production}.
\newblock \bibinfo{journal}{\emph{ACM Transactions on Graphics (TOG)}}
  \bibinfo{volume}{40}, \bibinfo{number}{6} (\bibinfo{year}{2021}),
  \bibinfo{pages}{1--18}.
\newblock


\bibitem[Paysan et~al\mbox{.}(2009)]%
        {paysan2009}
\bibfield{author}{\bibinfo{person}{Pascal Paysan}, \bibinfo{person}{Reinhard
  Knothe}, \bibinfo{person}{Brian Amberg}, \bibinfo{person}{Sami Romdhani},
  {and} \bibinfo{person}{Thomas Vetter}.} \bibinfo{year}{2009}\natexlab{}.
\newblock \showarticletitle{A 3D Face Model for Pose and Illumination Invariant
  Face Recognition}. In \bibinfo{booktitle}{\emph{2009 Sixth IEEE International
  Conference on Advanced Video and Signal Based Surveillance}}.
  \bibinfo{pages}{296--301}.
\newblock
\urldef\tempurl%
\url{https://doi.org/10.1109/AVSS.2009.58}
\showDOI{\tempurl}


\bibitem[Qi et~al\mbox{.}(2017)]%
        {qi2017pointnet}
\bibfield{author}{\bibinfo{person}{Charles~R Qi}, \bibinfo{person}{Hao Su},
  \bibinfo{person}{Kaichun Mo}, {and} \bibinfo{person}{Leonidas~J Guibas}.}
  \bibinfo{year}{2017}\natexlab{}.
\newblock \showarticletitle{Pointnet: Deep learning on point sets for 3d
  classification and segmentation}. In \bibinfo{booktitle}{\emph{Proceedings of
  the IEEE conference on computer vision and pattern recognition}}.
  \bibinfo{pages}{652--660}.
\newblock


\bibitem[Ranjan et~al\mbox{.}(2018)]%
        {ranjan2018generating}
\bibfield{author}{\bibinfo{person}{Anurag Ranjan}, \bibinfo{person}{Timo
  Bolkart}, \bibinfo{person}{Soubhik Sanyal}, {and} \bibinfo{person}{Michael~J
  Black}.} \bibinfo{year}{2018}\natexlab{}.
\newblock \showarticletitle{Generating 3D faces using convolutional mesh
  autoencoders}. In \bibinfo{booktitle}{\emph{Proceedings of the European
  conference on computer vision (ECCV)}}. \bibinfo{pages}{704--720}.
\newblock


\bibitem[Ravi et~al\mbox{.}(2020)]%
        {ravi2020pytorch3d}
\bibfield{author}{\bibinfo{person}{Nikhila Ravi}, \bibinfo{person}{Jeremy
  Reizenstein}, \bibinfo{person}{David Novotny}, \bibinfo{person}{Taylor
  Gordon}, \bibinfo{person}{Wan-Yen Lo}, \bibinfo{person}{Justin Johnson},
  {and} \bibinfo{person}{Georgia Gkioxari}.} \bibinfo{year}{2020}\natexlab{}.
\newblock \showarticletitle{Accelerating 3D Deep Learning with PyTorch3D}.
\newblock \bibinfo{journal}{\emph{arXiv:2007.08501}} (\bibinfo{year}{2020}).
\newblock


\bibitem[Roberts et~al\mbox{.}(2021)]%
        {deformationsurvey}
\bibfield{author}{\bibinfo{person}{Richard~A. Roberts},
  \bibinfo{person}{Rafael~Kuffner {dos Anjos}}, \bibinfo{person}{Akinobu
  Maejima}, {and} \bibinfo{person}{Ken Anjyo}.}
  \bibinfo{year}{2021}\natexlab{}.
\newblock \showarticletitle{Deformation transfer survey}.
\newblock \bibinfo{journal}{\emph{Computers Graphics}} (\bibinfo{year}{2021}).
\newblock
\showISSN{0097-8493}
\urldef\tempurl%
\url{https://doi.org/10.1016/j.cag.2020.10.004}
\showDOI{\tempurl}


\bibitem[Seol et~al\mbox{.}(2011)]%
        {seol2011artist}
\bibfield{author}{\bibinfo{person}{Yeongho Seol}, \bibinfo{person}{Jaewoo Seo},
  \bibinfo{person}{Paul~Hyunjin Kim}, \bibinfo{person}{John~P Lewis}, {and}
  \bibinfo{person}{Junyong Noh}.} \bibinfo{year}{2011}\natexlab{}.
\newblock \showarticletitle{Artist friendly facial animation retargeting}.
\newblock \bibinfo{journal}{\emph{ACM Transactions on Graphics (TOG)}}
  \bibinfo{volume}{30}, \bibinfo{number}{6} (\bibinfo{year}{2011}),
  \bibinfo{pages}{1--10}.
\newblock


\bibitem[Sharp et~al\mbox{.}(2022)]%
        {sharp2022diffusionnet}
\bibfield{author}{\bibinfo{person}{Nicholas Sharp}, \bibinfo{person}{Souhaib
  Attaiki}, \bibinfo{person}{Keenan Crane}, {and} \bibinfo{person}{Maks
  Ovsjanikov}.} \bibinfo{year}{2022}\natexlab{}.
\newblock \showarticletitle{Diffusionnet: Discretization agnostic learning on
  surfaces}.
\newblock \bibinfo{journal}{\emph{ACM Transactions on Graphics (TOG)}}
  \bibinfo{volume}{41}, \bibinfo{number}{3} (\bibinfo{year}{2022}),
  \bibinfo{pages}{1--16}.
\newblock


\bibitem[Song et~al\mbox{.}(2020)]%
        {song2020accurate}
\bibfield{author}{\bibinfo{person}{Steven~L. Song}, \bibinfo{person}{Weiqi
  Shi}, {and} \bibinfo{person}{Michael Reed}.} \bibinfo{year}{2020}\natexlab{}.
\newblock \showarticletitle{Accurate Face Rig Approximation with Deep
  Differential Subspace Reconstruction}.
\newblock \bibinfo{journal}{\emph{ACM Trans. Graph.}} \bibinfo{volume}{39},
  \bibinfo{number}{4}, Article \bibinfo{articleno}{34} (\bibinfo{date}{aug}
  \bibinfo{year}{2020}), \bibinfo{numpages}{12}~pages.
\newblock
\showISSN{0730-0301}
\urldef\tempurl%
\url{https://doi.org/10.1145/3386569.3392491}
\showDOI{\tempurl}


\bibitem[Sumner and Popovi{\'c}(2004)]%
        {sumner2004deformation}
\bibfield{author}{\bibinfo{person}{Robert~W Sumner} {and}
  \bibinfo{person}{Jovan Popovi{\'c}}.} \bibinfo{year}{2004}\natexlab{}.
\newblock \showarticletitle{Deformation transfer for triangle meshes}.
\newblock \bibinfo{journal}{\emph{ACM Transactions on graphics (TOG)}}
  \bibinfo{volume}{23}, \bibinfo{number}{3} (\bibinfo{year}{2004}),
  \bibinfo{pages}{399--405}.
\newblock


\bibitem[Tan et~al\mbox{.}(2018)]%
        {Tan_2018_CVPR}
\bibfield{author}{\bibinfo{person}{Qingyang Tan}, \bibinfo{person}{Lin Gao},
  \bibinfo{person}{Yu-Kun Lai}, {and} \bibinfo{person}{Shihong Xia}.}
  \bibinfo{year}{2018}\natexlab{}.
\newblock \showarticletitle{Variational Autoencoders for Deforming 3D Mesh
  Models}. In \bibinfo{booktitle}{\emph{Proceedings of the IEEE Conference on
  Computer Vision and Pattern Recognition (CVPR)}}.
\newblock


\bibitem[Tewari et~al\mbox{.}(2017)]%
        {tewari2017mofa}
\bibfield{author}{\bibinfo{person}{Ayush Tewari}, \bibinfo{person}{Michael
  Zollhofer}, \bibinfo{person}{Hyeongwoo Kim}, \bibinfo{person}{Pablo Garrido},
  \bibinfo{person}{Florian Bernard}, \bibinfo{person}{Patrick Perez}, {and}
  \bibinfo{person}{Christian Theobalt}.} \bibinfo{year}{2017}\natexlab{}.
\newblock \showarticletitle{Mofa: Model-based deep convolutional face
  autoencoder for unsupervised monocular reconstruction}. In
  \bibinfo{booktitle}{\emph{Proceedings of the IEEE International Conference on
  Computer Vision Workshops}}. \bibinfo{pages}{1274--1283}.
\newblock


\bibitem[Verma et~al\mbox{.}(2018)]%
        {verma2018feastnet}
\bibfield{author}{\bibinfo{person}{Nitika Verma}, \bibinfo{person}{Edmond
  Boyer}, {and} \bibinfo{person}{Jakob Verbeek}.}
  \bibinfo{year}{2018}\natexlab{}.
\newblock \showarticletitle{Feastnet: Feature-steered graph convolutions for 3d
  shape analysis}. In \bibinfo{booktitle}{\emph{Proceedings of the IEEE
  conference on computer vision and pattern recognition}}.
  \bibinfo{pages}{2598--2606}.
\newblock


\bibitem[Vesdapunt et~al\mbox{.}(2020)]%
        {vesdapunt2020jnr}
\bibfield{author}{\bibinfo{person}{Noranart Vesdapunt}, \bibinfo{person}{Mitch
  Rundle}, \bibinfo{person}{HsiangTao Wu}, {and} \bibinfo{person}{Baoyuan
  Wang}.} \bibinfo{year}{2020}\natexlab{}.
\newblock \showarticletitle{JNR: Joint-based neural rig representation for
  compact 3D face modeling}. In \bibinfo{booktitle}{\emph{Computer Vision--ECCV
  2020: 16th European Conference, Glasgow, UK, August 23--28, 2020,
  Proceedings, Part XVIII 16}}. Springer, \bibinfo{pages}{389--405}.
\newblock


\bibitem[Wu et~al\mbox{.}(2016)]%
        {wu2016anatomically}
\bibfield{author}{\bibinfo{person}{Chenglei Wu}, \bibinfo{person}{Derek
  Bradley}, \bibinfo{person}{Markus Gross}, {and} \bibinfo{person}{Thabo
  Beeler}.} \bibinfo{year}{2016}\natexlab{}.
\newblock \showarticletitle{An anatomically-constrained local deformation model
  for monocular face capture}.
\newblock \bibinfo{journal}{\emph{ACM transactions on graphics (TOG)}}
  \bibinfo{volume}{35}, \bibinfo{number}{4} (\bibinfo{year}{2016}),
  \bibinfo{pages}{1--12}.
\newblock


\bibitem[Wuu et~al\mbox{.}(2022)]%
        {wuu2022multiface}
\bibfield{author}{\bibinfo{person}{Cheng-hsin Wuu}, \bibinfo{person}{Ningyuan
  Zheng}, \bibinfo{person}{Scott Ardisson}, \bibinfo{person}{Rohan Bali},
  \bibinfo{person}{Danielle Belko}, \bibinfo{person}{Eric Brockmeyer},
  \bibinfo{person}{Lucas Evans}, \bibinfo{person}{Timothy Godisart},
  \bibinfo{person}{Hyowon Ha}, \bibinfo{person}{Alexander Hypes},
  \bibinfo{person}{Taylor Koska}, \bibinfo{person}{Steven Krenn},
  \bibinfo{person}{Stephen Lombardi}, \bibinfo{person}{Xiaomin Luo},
  \bibinfo{person}{Kevyn McPhail}, \bibinfo{person}{Laura Millerschoen},
  \bibinfo{person}{Michal Perdoch}, \bibinfo{person}{Mark Pitts},
  \bibinfo{person}{Alexander Richard}, \bibinfo{person}{Jason Saragih},
  \bibinfo{person}{Junko Saragih}, \bibinfo{person}{Takaaki Shiratori},
  \bibinfo{person}{Tomas Simon}, \bibinfo{person}{Matt Stewart},
  \bibinfo{person}{Autumn Trimble}, \bibinfo{person}{Xinshuo Weng},
  \bibinfo{person}{David Whitewolf}, \bibinfo{person}{Chenglei Wu},
  \bibinfo{person}{Shoou-I Yu}, {and} \bibinfo{person}{Yaser Sheikh}.}
  \bibinfo{year}{2022}\natexlab{}.
\newblock \showarticletitle{Multiface: A Dataset for Neural Face Rendering}. In
  \bibinfo{booktitle}{\emph{arXiv}}.
\newblock
\urldef\tempurl%
\url{https://doi.org/10.48550/ARXIV.2207.11243}
\showDOI{\tempurl}


\bibitem[Yang et~al\mbox{.}(2022)]%
        {yang2022implicit}
\bibfield{author}{\bibinfo{person}{Lingchen Yang}, \bibinfo{person}{Byungsoo
  Kim}, \bibinfo{person}{Gaspard Zoss}, \bibinfo{person}{Baran G\"{o}zc\"{u}},
  \bibinfo{person}{Markus Gross}, {and} \bibinfo{person}{Barbara Solenthaler}.}
  \bibinfo{year}{2022}\natexlab{}.
\newblock \showarticletitle{Implicit Neural Representation for Physics-Driven
  Actuated Soft Bodies}.
\newblock \bibinfo{journal}{\emph{ACM Trans. Graph.}} \bibinfo{volume}{41},
  \bibinfo{number}{4}, Article \bibinfo{articleno}{122} (\bibinfo{date}{jul}
  \bibinfo{year}{2022}), \bibinfo{numpages}{10}~pages.
\newblock
\showISSN{0730-0301}
\urldef\tempurl%
\url{https://doi.org/10.1145/3528223.3530156}
\showDOI{\tempurl}


\bibitem[Zhou et~al\mbox{.}(2020)]%
        {zhou2020fully}
\bibfield{author}{\bibinfo{person}{Yi Zhou}, \bibinfo{person}{Chenglei Wu},
  \bibinfo{person}{Zimo Li}, \bibinfo{person}{Chen Cao},
  \bibinfo{person}{Yuting Ye}, \bibinfo{person}{Jason Saragih},
  \bibinfo{person}{Hao Li}, {and} \bibinfo{person}{Yaser Sheikh}.}
  \bibinfo{year}{2020}\natexlab{}.
\newblock \showarticletitle{Fully convolutional mesh autoencoder using
  efficient spatially varying kernels}.
\newblock \bibinfo{journal}{\emph{Advances in Neural Information Processing
  Systems}}  \bibinfo{volume}{33} (\bibinfo{year}{2020}),
  \bibinfo{pages}{9251--9262}.
\newblock


\end{thebibliography}
\clearpage
\newpage

\clearpage
\appendix

\section{Implementation Details} 
    \new{We render the front views of $S$ and $T$ by Pytorch3D~\cite{ravi2020pytorch3d} with a fixed point light and a gray Lambert shading. The model is not sensitive to these settings as long as the whole face is shown in the image.}  

The rendered 256$\times$256 RGBD images feed into $CNN_{ST}$ which has four 2D convolution layers followed by a fully-connected layer to output $c_S$ and $c_T$. $\text{DN}_e$ and $\text{DN}_i$ have four and two diffusion layers, respectively.   $MLP_{dec}$ contains eight linear layers of 256 hidden dimensions, with ReLU activations in between.

\new{We train the model with $\lambda_v = 10$, $\lambda_g = 1$, $\lambda_n = 1$ and $\lambda_e = 0.1$. During the first training stage, we warm up the MLP by providing the ground truth $z_{FACS}$ in the first 100 epochs. And then train another 200 epochs with $z^*_{FACS}$ from the expression encoder. In the second stage, we train on both \emph{ICT-Real-AU} and \emph{Multiface} until convergence. We set the initial learning rate as 1e-4 and decrease it by a factor of 0.75 for every 100 epochs.}

\section{Data Augmentation}
    
Here we visualize the output meshes after data augmentation. 
\begin{figure}[H]
\centering
        \begin{subfigure}[b]{0.2\linewidth}
                \includegraphics[width=\linewidth]{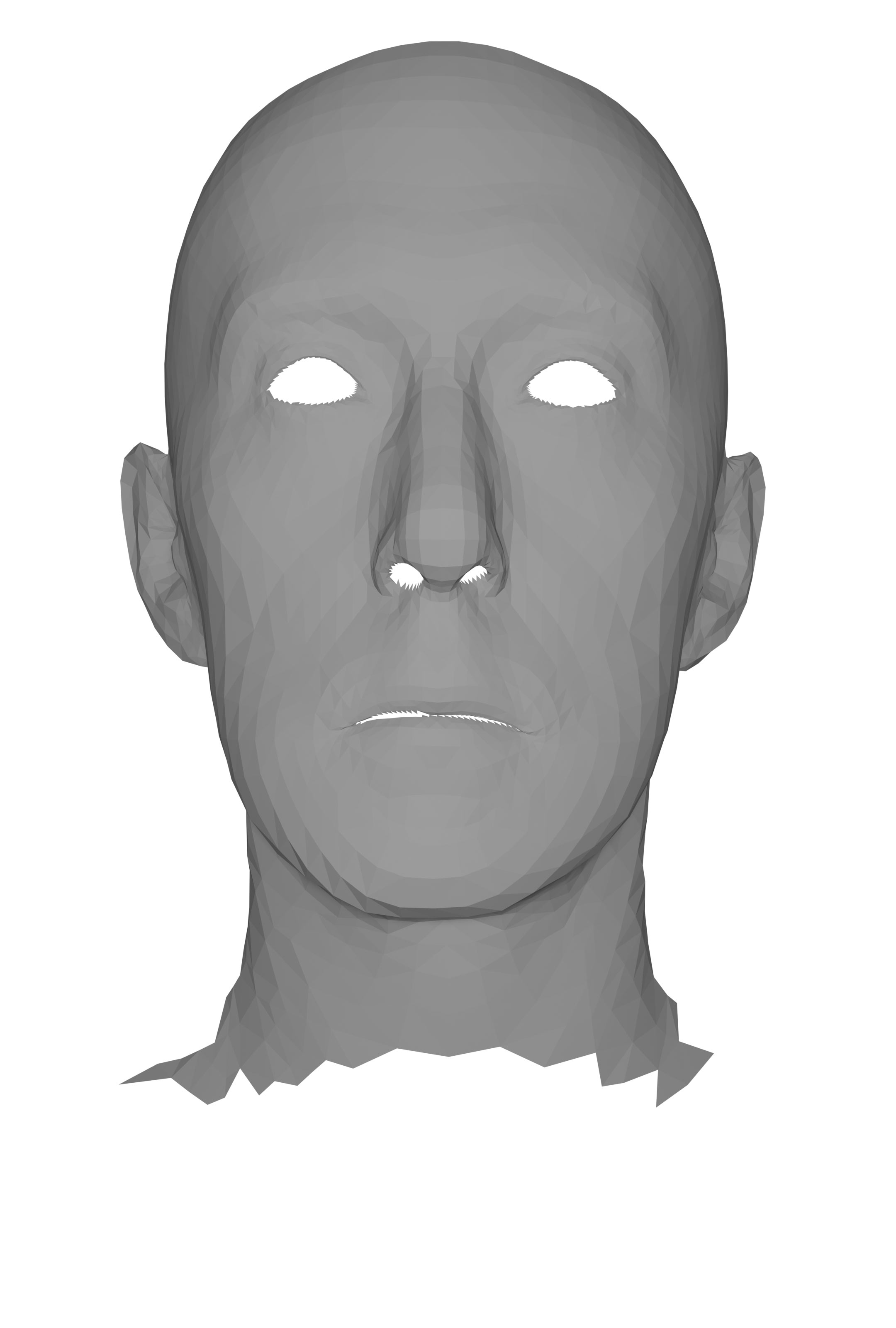}
                \caption{}
                \label{fig:ict_ss}
        \end{subfigure}%
        \begin{subfigure}[b]{0.2\linewidth}
                \includegraphics[width=\linewidth]{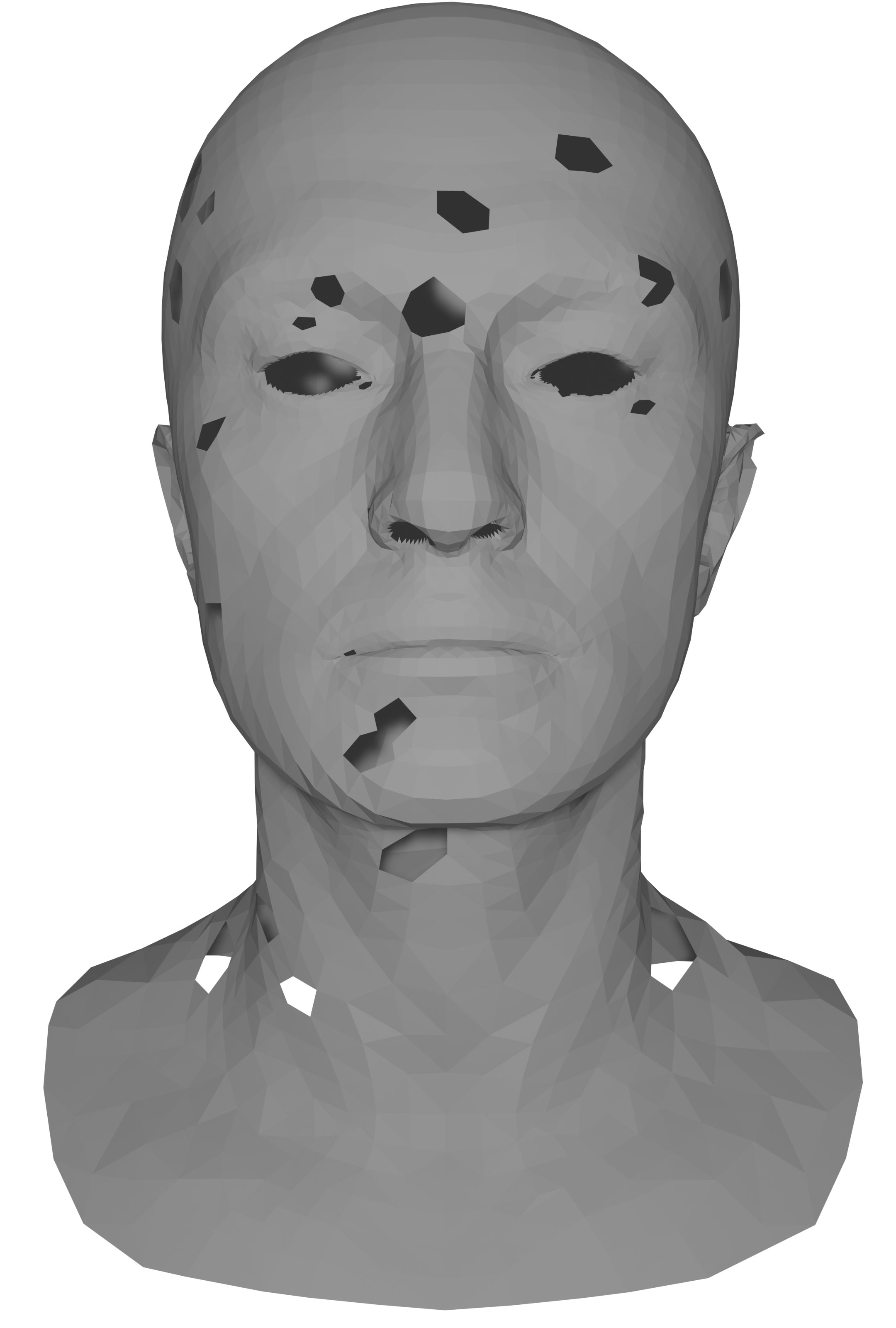}
                \caption{}
                \label{fig:ict_holes}
        \end{subfigure}%
        \begin{subfigure}[b]{0.2\linewidth}
                \includegraphics[width=\linewidth]{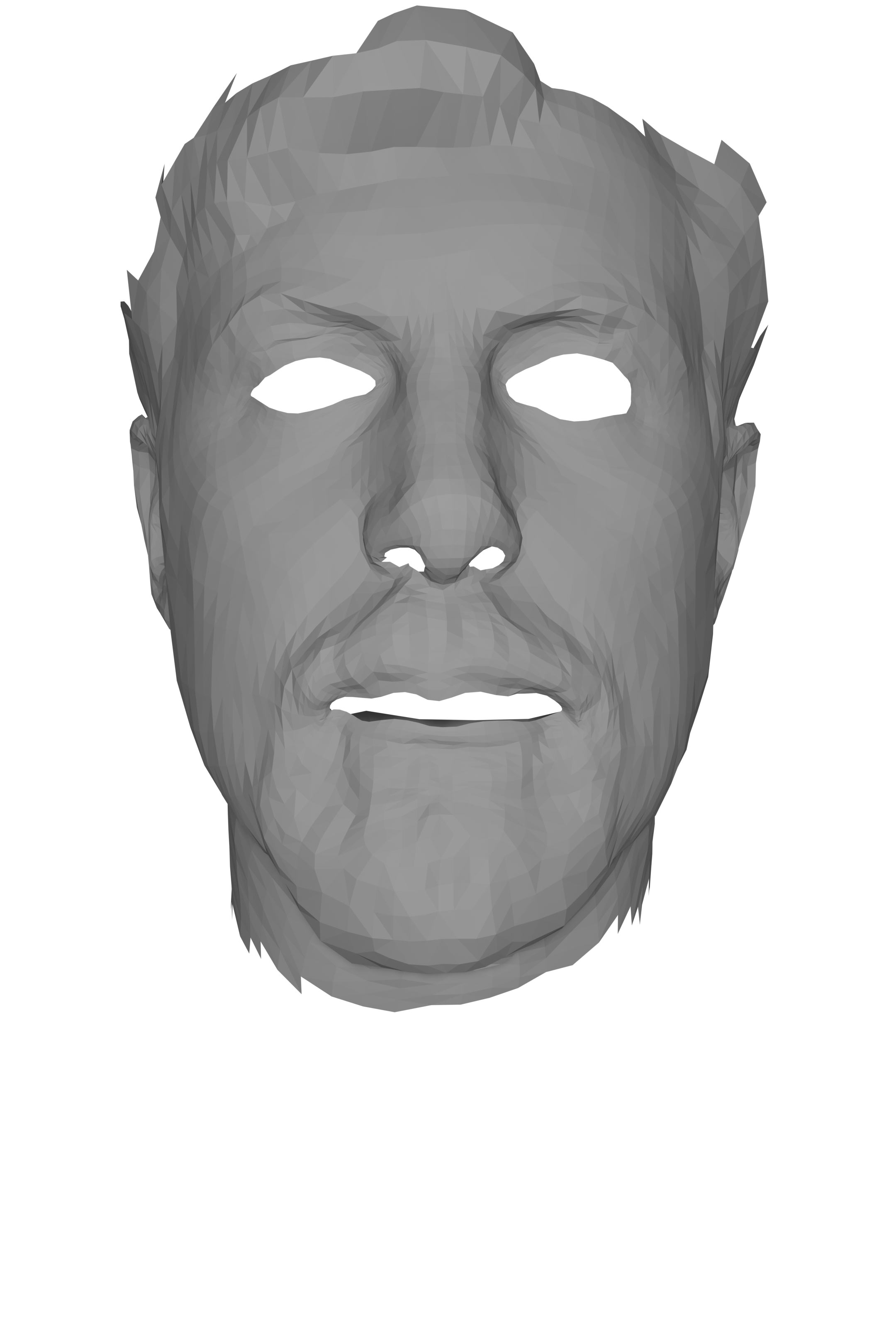}
                \caption{}
                \label{fig:mf_ss}
        \end{subfigure}%
        \begin{subfigure}[b]{0.2\linewidth}
                \includegraphics[width=\linewidth]{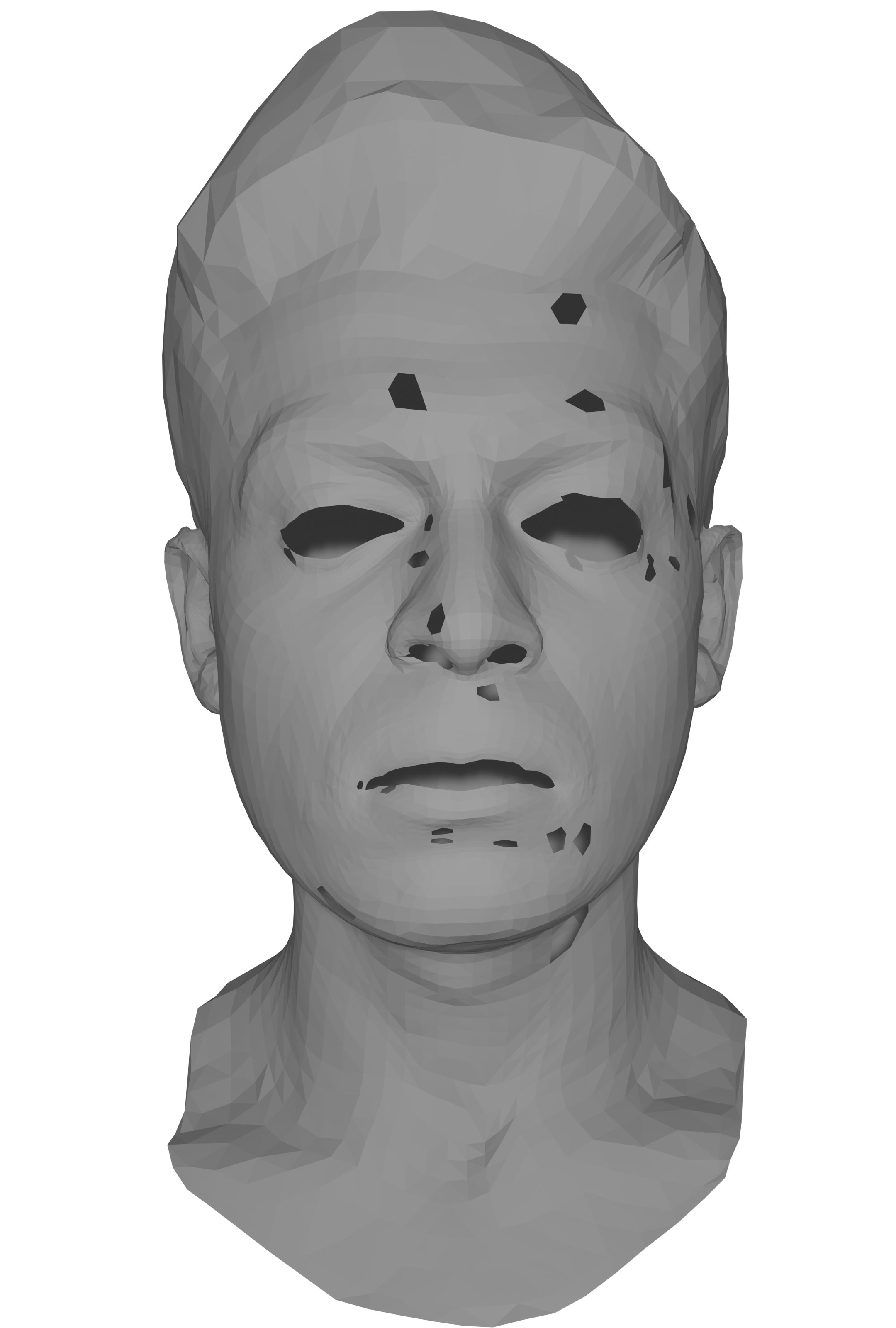}
                \caption{}
                \label{fig:mf_holes}
        \end{subfigure}
        \caption{\new{\textbf{Data augmentation scheme.} (a) and (c) show shift, scale, and applying a front-facial mask. (b)and (d) show holes cutting. (a), (b) are identities from the \emph{ICT} and (c), (d) are from \emph{Multiface}.}}
        \label{fig:augmentation}
\end{figure}
\section{Standardization}
    
\label{standardization} 
\begin{figure}[H]
    \centering
    \begin{subfigure}[b]{0.45\linewidth}
                \includegraphics[width=\linewidth]{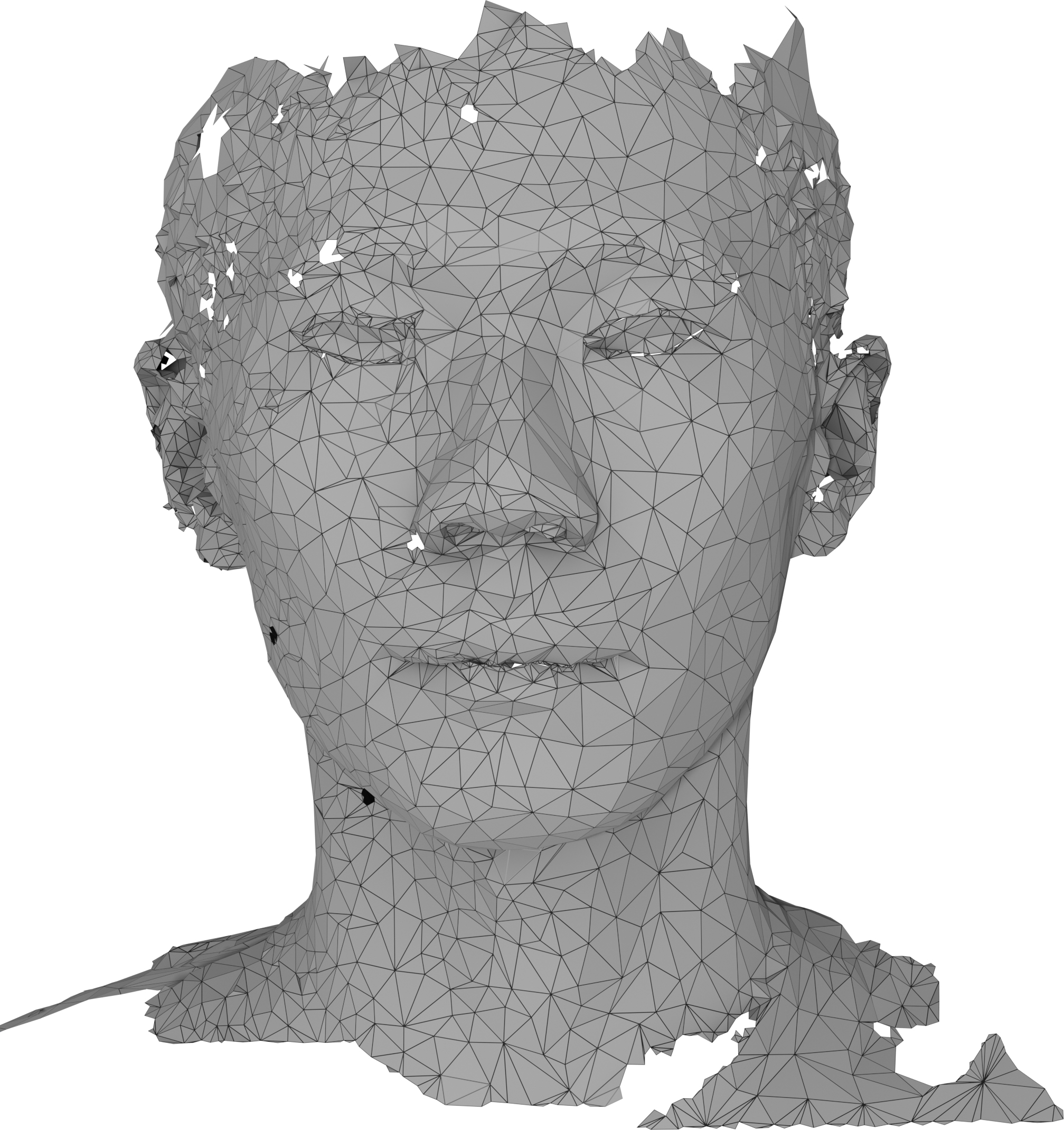}

                \label{fig:raw}
        \end{subfigure}%
    \begin{subfigure}[b]{0.45\linewidth}
                \includegraphics[width=\linewidth]{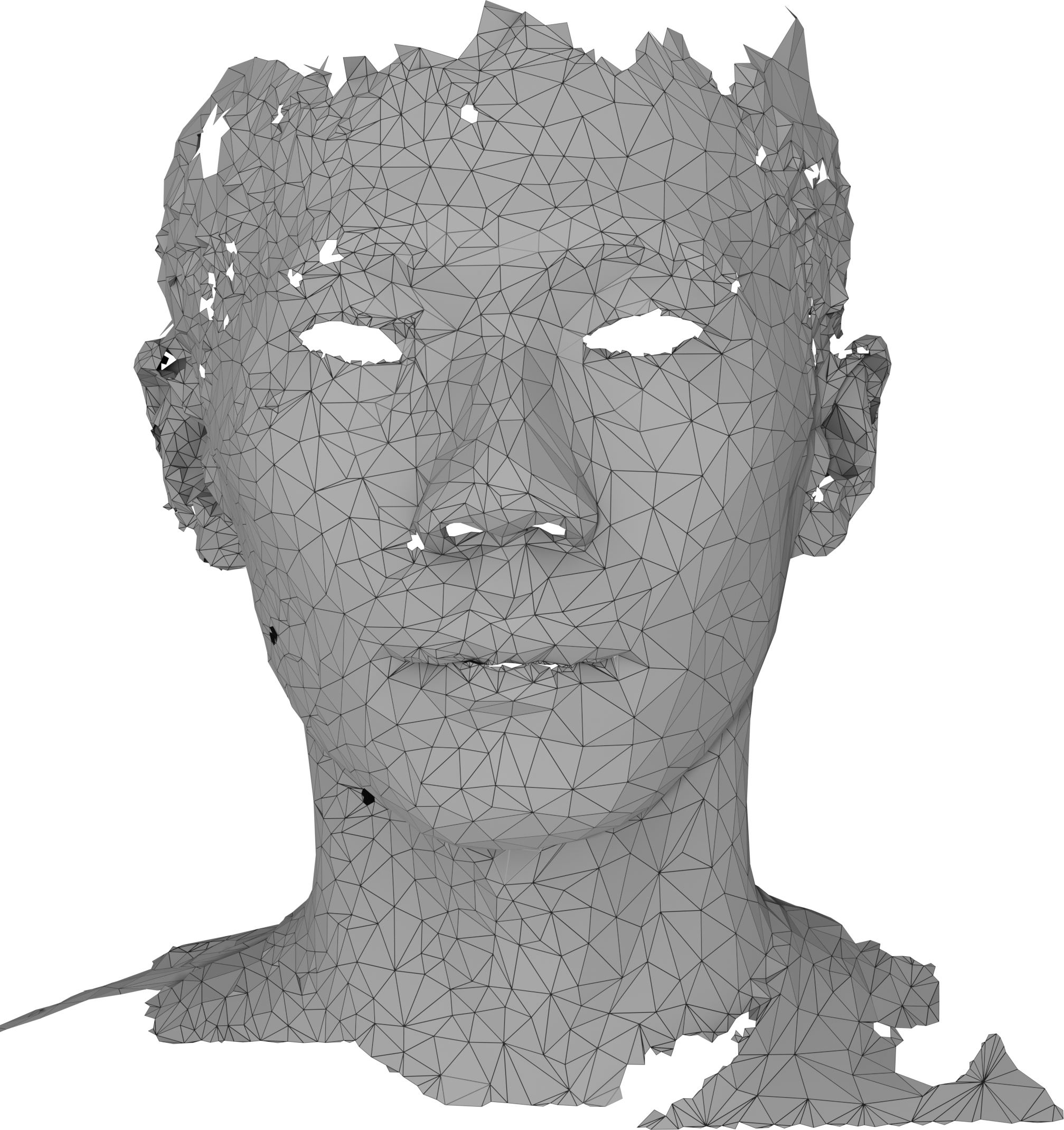}

                \label{fig:raw_process}
        \end{subfigure}%
        
    \caption{\new{\textbf{Left:} Raw Scan. \textbf{Right:} Raw Scan after standardization. For an in-the-wild mesh with connected eyes, nose, and mouth, we need to cut those areas to have the correct global solving of the deformation transfer process.}}
    \label{fig:raw_stand}
\end{figure}

\begin{figure}[H]
\centering
\includegraphics[width=0.8\linewidth]{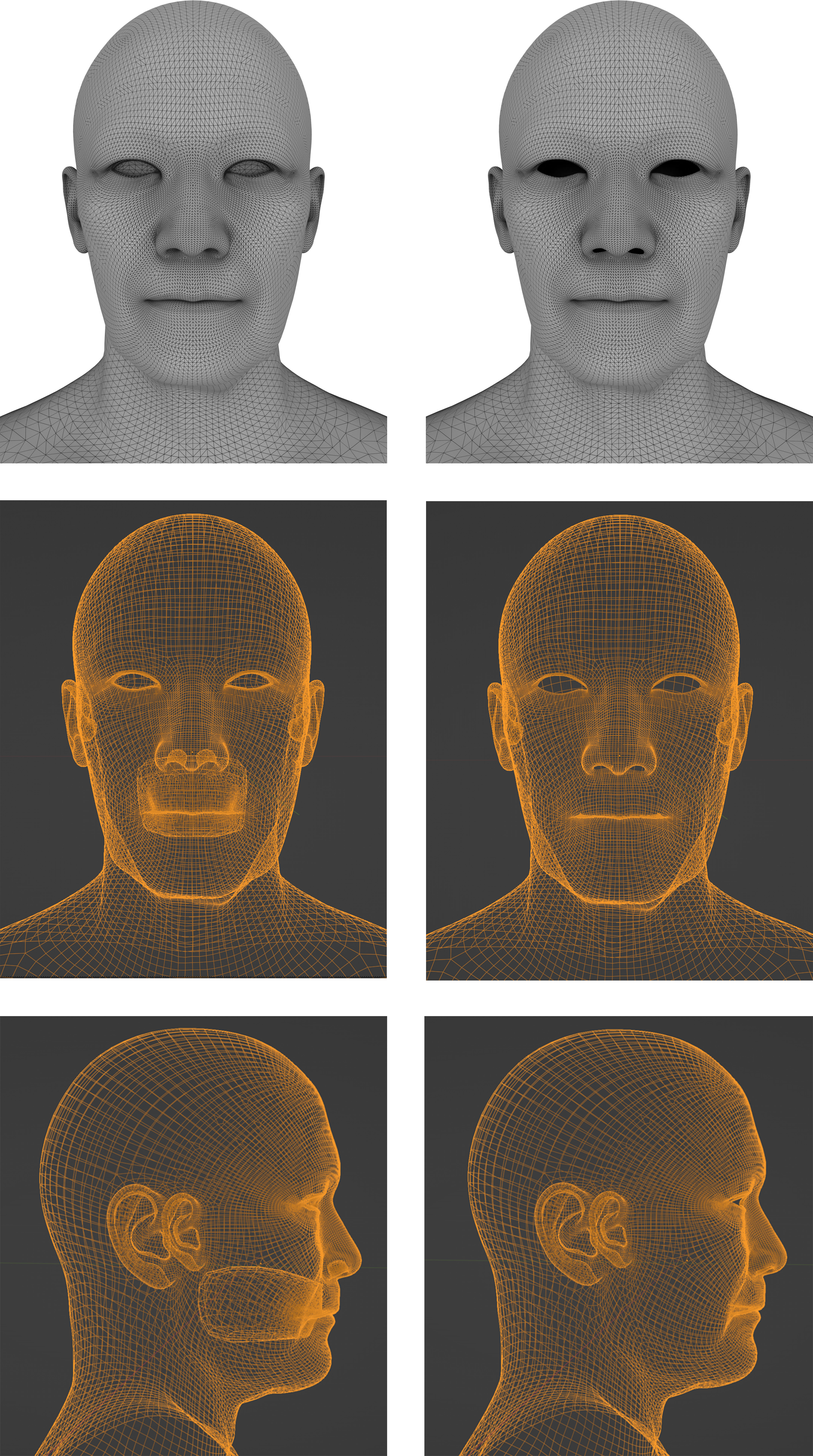}
\caption{\new{\textbf{Left:} The original \textit{MetaHuman} mesh. \textbf{Right:} The same mesh after standardization. We remove the inner mouth socket to make it consistent with other meshes during training.}}
\label{fig:metahuman_stand}
\end{figure}

\new{We give two examples of the data standardization treatment. Fig. \ref{fig:raw_stand} illustrates the process for an in-the-wild raw scan, while Fig. \ref{fig:metahuman_stand} shows the pipeline for an artist-created mesh.}

\end{document}